\documentclass{comnet}
\usepackage[T1]{fontenc} %% needef for \k in Was (\k{a} does not work in OT1) Also possible [T2A]
\usepackage{soul}
\usepackage{array}
\usepackage{subfigure}
\usepackage{graphicx}
\usepackage{amsmath,amssymb,amsfonts}
\usepackage{amsthm}
\usepackage{bm}
\usepackage{url}
\usepackage{xcolor}
\usepackage{enumitem}
\newcommand{\diag}{\mathop{\mathrm{diag}}}

\def\sign{\mathop{\rm sign}\nolimits}
\def\varnothing{\emptyset}
\newcommand{\x}[1]{} %\def\x#1{} %\x{} For in-line comments
\def\xy{\hspace{.07em}}          %+
\def\xxy{\hspace{.14em}}          %+
\def\xz{\hspace{-.07em}}         %+
\def\xx{X}
\def\yy{Y}
\def\zz{Z}
\def\cdc{,\ldots,}
\def\N{{\mathbb N}}
\def\R{{\mathbb R}}
\def\M{{\mathcal M}}
\def\PP{{\mathcal P}}

\def\a{\mbox{{\small$\alpha$}}}                 %{\alpha}
\def\be{\mbox{{\small$\beta$}}}
\def\ve{\a}           %!!{\varepsilon}
\def\ph{\varphi}
\def\ccdot{\!\cdot\!}
\def\phc{\ph}          %{\ph(\cdot)}
\def\phcc{\ph(\cdot)}
\def\phP{\ph^{}_\PP}
\def\phPc{\phP}        %{\phP(\cdot)}
\def\uti{\omega}

\def\sssm{\!\smallsetminus\!}
\def\Closeness{{\it Closeness\/}}
\def\Connectivity{{\it Connectivity\/}}
\def\Betweenness{{\it Betweenness\/}}

\def\PageRank{{\it PageRank\/}}
\def\Degree{{\it Degree\/}}
\def\Seeley{{\it Seeley\/}}
\def\PageRankCentrality{{\it PageRank centrality\/}}
\def\EstradaSubgraph{{\it Estrada subgraph\/}}
\def\Walk{{\it Walk\/}}
\def\Katzc{{\it Katz\/}} %!{{\it Walk\/}}
\def\KatzcCentrality{{\it Katz centrality\/}} %!{{\it Walk centrality\/}}
\def\Forest{{\it Forest\/}}
\def\Heat{{\it Heat\/}}
\def\Bonacich{{\it Bonacich\/}}
\def\BonacichCentrality{{\it Bonacich centrality\/}}
\def\Eigenvector{{\it Eigenvector\/}}
\def\EigenvectorCentrality{{\it Eigenvector centrality\/}}
\def\Communicability{{\it Communicability\/}}
\def\GeneralizedDegree{{\it Generalized degree\/}}
\def\GeneralizedDegreeCentrality{{\it Generalized degree centrality\/}}
\def\TotalCommunicability{{\it Total communicability\/}}
\def\TotalWalk{{\it Total walk\/}}
\def\Eccentricity{{\it Eccentricity\/}}
\def\amonotonic{{a supporting}}     %\hl{an increasing}
\def\monotonic{supporting}          %increasing

\def\1n{1\cdc n}
\newcommand{\eq}[2]{\begin{equation}\label{#1}#2\end{equation}}
\def\eqs*#1{\begin{eqnarray*}#1\end{eqnarray*}}
\def\eqss#1{\begin{eqnarray}#1\end{eqnarray}}
\def\beq#1{\begin{equation}\label{#1}}
\def\eeq{\end{equation}}
\definecolor{ggreen}{rgb}{0,.27,0}
\renewcommand{\hl}[1]{{#1}}    % highlighting from soul package is neutralized
\newcommand{\aB}[1]{{\color{black}#1}}    % comments of Author B
\newcommand{\ax}[2]{#2}
\def\Comm{{\operatorname{Comm}}}    %\mathrm
\def\Katz{{\operatorname{Walk}}}
\def\Hea{\operatorname{Heat}}
\def\For{\operatorname{For}} %{regL}
\def\Up#1{\vspace{-#1em}}                                       %+

\def\succE{\succcurlyeq}
\def\To{\Rightarrow}
\def\ToTo{\Leftrightarrow}
\def\painfty{\bm{+\infty}}
\def\mainfty{\bm{-\infty}}

\def\Equivalence{Equivalence}
\def\Ee{A{\xz}E}   %{\r{E}}
\def\fPRa{f^{\operatorname{PR}_{\alpha}\!}}

\def\pathcondition{path centripetal condition}
\def\Axiom{axiom}

\setlist[itemize]{noitemsep, topsep=0pt}

\begin{document}
\title{\vspace{0.0em}Selection of Centrality Measures Using Self-consistency and Bridge Axioms%  \x{ by Means of}
}

\shorttitle{Selection of Centrality Measures Using Self-consistency and Bridge Axioms} %%%for recto running head
\shortauthorlist{Pavel Chebotarev} %%% for verso running head

\author{%%%% First author details
\name{Pavel Chebotarev}
\address{Department of Mathematics, Technion--Israel Institute of Technology, Haifa, 3200003 Israel and\\
         Kharkevich Institute for Information Transmission Problems, RAS, Bol'shoi Karetnyi per., 19, Moscow, 127051 Russia\email{pavel4e@technion.ac.il}} %pavel4e@gmail.com
}

\maketitle

%\vspace{-2em}
\begin{abstract}
{We consider several families of network centrality measures induced by graph kernels, which \ax{contain}{include} some well-known measures and many new ones.
The Self-consistency and Bridge axioms, which appeared earlier in the literature, are closely related to certain kernels and one of the families.
We obtain a necessary and sufficient condition for Self-consistency, a sufficient condition for the Bridge axiom, indicate specific measures that satisfy these axioms, and show that under some additional conditions they are incompatible.
\x{It is also shown that }PageRank centrality applied to undirected networks violates most conditions under study and has a property that according to some authors is ``hard to imagine''\x{ hardly imaginable} for a centrality measure.
We explain\x{ the reason for} this phenomenon.
Adopting the Self-consistency or Bridge axiom leads to a drastic reduction in survey time in the culling method designed to select the most appropriate centrality measures.}
{undirected network; centrality measure; axiomatic approach; self-consistency; bridge axiom; culling method; PageRank.}
\\
2000 Math Subject Classification:
05C50, %Graphs and matrices
05C82, %Small world graphs, complex networks (graph-theoretic aspects)
15A16, %Matrix exponential and similar functions of matrices
65F60, %Numerical computation of matrix exponential and similar matrix functions
91D30, %Social networks; opinion dynamics
94C15  %Applications of graph theory to circuits and networks
\end{abstract}

\section{Introduction and related work}
\label{s:intro}

The number of network centrality measures studied in the literature exceeds 400~\cite{centiserver} and\x{ many} new measures appear every year.
This diversity needs to be structured.
The main means of structuring it is to establish a correspondence between the measures and their properties, some of which can be treated as\x{ considered as normative conditions or} axioms.
The purpose of this paper is to advance this work by studying two natural axiomatic conditions, namely, the Self-consistency and Bridge axioms, which are closely related to some kernels and a certain class of kernel-based centrality measures. We establish a sufficient condition for the Bridge axiom, a necessary and sufficient condition for Self-consistency, and indicate centralities, some of which are well known and others are new, that satisfy these axioms.

Very often, centrality is identified with structural importance or influence~\cite{Preston70,WassermanFaust94,Geisberger08,RoyTredan10SharpCentr,Lockhart16edge,Mavrodiev21bats}. However, there are kinds\x{ concepts} of importance that are not reducible to centrality. Say, in a chain of moving people modeled by a path graph, the most important\x{ agents} actors may be the leader and the trailer, i.e., the least central end elements of the chain, while the central elements\x{ of such a chain} may not be of particular importance. \aB{For another case where the key players are peripheral rather than central, see~\cite{Tyloo19oscill}.} Thus, importance of network nodes\x{ in a network} is a broader concept\x{ more diverse and general notion} than centrality.%the importance of nodes in networks is not necessarily manifested through centrality.

Anyway, each point centrality index measures some {\em structural capital\/} (cf. \cite{Jackson20Capital}) of the nodes.
It turns out that the\x{ kinds} types of capital accounted for by the centralities that satisfy the Bridge axiom on the one hand and by centralities satisfying\x{ the conjunction of} Self-consistency and Monotonicity on the other hand are fundamentally\x{ significantly} different, so that\x{ and therefore} these conditions are incompatible, \x{whenever}provided that the \Equivalence\ axiom is assumed\x{ to hold (obeyed)}. Similarly, the Bridge axiom is incompatible with another positive responsiveness axiom called Transit monotonicity.
In terms of another taxonomy \cite{BlochJackson23-}, these axioms are related to different types of {\em nodal statistics}: path-based for the Bridge axiom and walk-based for Self-consistency.

The axiomatic approach commences when we transition from examining the properties of specific objects to examining the properties in and of themselves. At this stage, some of the properties become requirements, and the objective becomes the search for objects that satisfy these requirements or their combinations.

{Axiomatic study of\x{ point} centrality measures was initiated by Sabidussi} \cite{sabidussi1966centrality} (whose starting point was the analysis of the \Degree\ and\x{ Bavelas's} \Closeness\ centralities) and continued by Nieminen \cite{Nieminen73Directed,nieminen1974centrality} and\x{ then many} others. A good review of early work is~\cite{boldi2014axioms}. \hl{Among the papers published later or not mentioned in} \cite{boldi2014axioms}, interest\-ing axioms and results can be found, in particular, in \cite{palacios2004measurement,altman2008axiomatic,garg2009axiomatic,kitti2016axioms,boldi2017rank,csato2017measuring,dequiedt2017local,bandyopadhyay2018generic,schoch2018centrality,%
skibski2018axioms,skibski2019attachment,Boldi22Monotonicity-,BlochJackson23-,Skibski23Closeness,WasSkibski23Pagerank}. %
%altman2005ranking % altman2005ranking

{In many cases, the object of axiomatization is the procedure for assigning the most central node in a network} \cite{holzman1990axiomatic,vohra1996axiomatic,monsuur2004centers}, which implicitly characterizes the corresponding centrality measure.

{In this paper, we focus on the Self-consistency and Bridge axioms, both of which are ordinal in nature. The first states that centrality measures must reward nodes for having neighbors with high centrality. The second one requires the measures
to \hl{always} reward the bridge end-node for the \hl{larger} size of the component it belongs to after removing the bridge.}

{In the case of directed graphs that express paired comparisons, Self-consistency appeared in} \cite{CheSha97a,CheSha98AOR,CheSha99,DiazHendrickxLohmann13SCW,Csato19ImposPC,Csato19ImposT}; in~\cite{Csato19Journal} it was applied to reference networks; in the case of undirected graphs, we find it in \cite{bandyopadhyay2018generic} under the name of Structural consistency. It strengthens Preservation of neighborhood-inclusion~\cite{schoch2018centrality}, whose directed graph version goes back to Preservation of cover relation~\cite{miller1980new}. %bandyopadhyay2017generic-,

{The Bridge axiom was proposed in}~\cite{skibski2018axioms} to characterize the \Closeness\ centrality; in \cite{Skibski23Closeness} it was strengthened to Majority comparison; in \cite{Khmelnitskaya23}, the Ratio property, another strengthening of the Bridge axiom, was presented.

{In addition to the Self-consistency and Bridge axioms, the class of positive responsiveness axioms contains a number of monotonicity conditions, the study of which has a long history. We use a Monotonicity axiom similar to those proposed in} \cite{csato2017measuring,Boldi22Monotonicity-} and in \cite{Che94,ChienDwork04link,boldi2017rank} for directed graphs. Transit monotonicity is a natural strengthening of Monotonicity, which is\x{ shown to be} incompatible with the Bridge axiom.
%A necessary and sufficient condition for Self-consistency we prove involves some result on the extension of utility functions.

{The Bridge and Self-consistency axioms are satisfied by certain kernel-based centralities. In this paper, we present several new classes of such centralities and illustrate the difference between their representatives. The corresponding kernels and some related centralities have been introduced by\x{ many} different authors; the references to their \hl{works} are provided in} Section~\ref{s_NewMes}. %\x{ papers}

\PageRank\ is a centrality measure that has attracted a lot of attention. \x{In this paper, }We show that, when applied to undirected networks, it violates\x{ does not satisfy the!} most of the conditions under consideration and explain this phenomenon.

The paper is organized as follows.
After introducing the basic notation in Section~\ref{s:notat}, in Section~\ref{s_NewMes} we consider several families of centralities defined using\x{ related to (associated with)} graph kernels. In Section~\ref{s:axioms}, the Bridge and Self-consistency axioms are introduced. %presented. discussed.
Section~\ref{s:Bridge} presents a sufficient condition for the Bridge axiom as well as a number of measures that satisfy it. In Section~\ref{s:Self-c}, we prove a necessary and sufficient condition for Self-consistency and present centralities that meet\x{ satisfy} it.
Section~\ref{s:intuit} discusses simple general requirements for centrality measures.\x{, simple general requirements to centrality measures are discussed.} In Section~\ref{s:Monot}, we consider the axioms of Monotonicity and Transit monotonicity and prove that the addition of them is sufficient to ensure the properties of Section~\ref{s:intuit} and to form conditions incompatible with the Bridge axiom. The final\x{concluding} Section~\ref{s:Discuss} contains some interpretations of the results obtained and discusses some features of \PageRank.
%\section{Related work}
%\label{s_Related}

\section{Notation}
\label{s:notat}

Let $G=(V,E)$ be an undirected unweighted graph with node set $V=V(G)$ and edge set $E=E(G).$
The {\em order\/} of $G$ is $|V|=n.$
Graph nodes will be labeled with\x{ letters} $u,v,w,u_i,v_i$, etc., numbers $0, 1, 2,\ldots,$ or names: Medici, Pazzi, etc.
We consider graphs with $n>1,$ without loops and multiple edges.
Since some centrality measures under study are applicable only to connected graphs, we\x{ restrict our attention} confine ourselves to them.

Nodes $u$ and $v$ of $G$ are {\em neighbors\/} iff $\{u,v\}\in E(G).$ Let $N_u$ denote the set of neighbors of node~$u.$ %in~$G.$

The {\em adjacency matrix\/} of $G$ is denoted by $A=A(G)=(a_{uv})_{n\times n}$: $a_{uv}=1$ when $u$ and $v$ are neighbors and $a_{uv}=0,$ otherwise. Neighborhood is a symmetric relation, so $A$ is symmetric\x{$A^T=A$}.
Let $\rho(A)$ be the spectral radius of~$A.$

The \emph{degree\/} $d_u$ of a node $u$ is the number of\x{ edges incident to} neighbors of~$u$: $d_u=|N_u|.$
The vector of node degrees is $\bm d=(d_1\cdc d_n)^T=A\bm 1,$ where $\bm1=(1\cdc 1)^T.$
A \emph{leaf\/} is a node that has exactly one neighbor.
Nodes $u$ and $v$ are \emph{equivalent\/} in $G$ if there exists an automorphism\footnote{An automorphism of $G$ is a permutation $\sigma$ of $V(G)$ such that for any $u,v\in V(G),$ $\{u,v\}\in E(G)\ToTo$\x{ and only if} $\{\sigma(u),\sigma(v)\}\in E(G)$.}
of $G$ that takes $u$ to $v$\xy; in this case we write $u\sim v.$

The {\em Laplacian matrix\/} of $G$ is
\beq{e:L}
L=\diag(A\bm1)-A, %\;\;\mbox{where}\;\;  \bm1=(1\cdc 1)^T.
\eeq
where $\diag(\bm x)$ is the diagonal matrix with vector $\bm x$ on the diagonal.
%
%The \emph{shortest path distance\/} between two nodes is the length of a shortest path between them.

The union $G=G_1\cup G_2$ of arbitrary graphs $G_1$ and $G_2$\x{ (not necessarily disjoint)} is defined by: $V(G)=V(G_1)\cup V(G_2)$ and $E(G)=E(G_1)\cup E(G_2)$.

Given a graph $G,$ %{$G=(V,E)$ with set of nodes $V=V(G)$ and set of edges $E=E(G)$}
a {\em centrality measure\/} (or \emph{centrality\/}; sometimes, {\em point centrality\/}) $f\!:V(G)\to\R$ associates a real number $f(v)$ with each node $v\in V(G)$.
For every $v\in V(G)$, $f(v)$ depends only on the graph structure and the position of $v$ in it~\cite{WassermanFaust94}.
\x{Thus, $f$ depends on $G,$ However, }For simplicity, we reflect the connection of $f$ with $G$ in the notation only when two or more graphs are considered simultaneously.
%In most cases $G$ is fixed, and when it is not, we explicitly specify the graph to which centrality applies.
%\x{Formally, for a fixed graph $G$, a centrality on $G$ is a function $f\!: V(G)\to \R_+\cup\{0\}.$}
%It associates a non-negative real number $f(v)$ with every node $v\in V(G)$ based only on the graph structure~\cite{WassermanFaust94}. %,RoyTredan10SharpCentr
%\item A \emph{centrality measure}, or \emph{centrality} on $G$ is a function $f: V(G)\to \R_+\cup\{0\}$.
Various conceptions of centrality are quite diverse.
In this regard, there is no generally accepted definition of centrality that would semantically distinguish it from other types of point structural measures. A few simple attempts\x{ approaches} to make such a distinction are discussed in Section~\ref{s:intuit}.
%The number $f_i(v)$ is interpreted as the \emph{centrality\/} of $v$ in $G$ assigned by~$f_i$: the more $f_i(v),$ the more central $v$ is considered. In what follows, $f$ denotes any centrality measure.
%When there is a need to specify $G$ explicitly, we write~$f^G.$ % will

When a centrality measure $f$ on $G$ is fixed, we will write $u\succ v$ and $u\simeq v$ %$u\succeq v,$ and $u\cong v$
as\x{ a short version} abbreviations for $f(u)>f(v)$ and $f(u)>f(v)$, respectively. %$f(u)\ge f(v),$ and $f(u)=f(v)$, respectively.
Moreover, if, for instance\x{example}, $V=\{1\cdc 7\},$ then $(\{1,6\},\{2,3,4\},5,7)$ is an example of\x{ \emph{ranking of nodes by centrality} (or,} \emph{centrality ranking\/} of nodes $1$ to $7$\x{, namely, this means that} in which $f(1)=f(6)>f(2)=f(3)=f(4)>f(5)>f(7).$
%Among the centralities under considerations, there are measures applicable only to connected graphs, so we restrict our consideration to such graphs; for nondegeneracy, we assume $|V(G)|>1$.

\section{Centrality measures induced by graph kernels} %proximity measures and distances
\label{s_NewMes}

In this section, we present\x{introduce,consider} several families of centrality measures.

Let $d(u,v)$ be the \emph{shortest path distance\/} \cite{BuckleyHarary90} between nodes $u$ and $v$ in a graph\x{ $G$}, i.e., the length of a shortest path between $u$ and~$v.$
Two popular\footnote{For example, in\x{ the recent study} \cite[\hl{p.\,1}]{ali2020revisit}, the authors come to the conclusion that in the infection source identification problem, ``a combination of eccentricity and closeness...\ generally performs better than several state-of-the-art source identification techniques, with higher accuracy and lower average hop error''. One more popular distance based measure is the {\em Harmonic closeness\/} $f(u)=\sum_{v\ne u}(d(u,v))^{-1}$.\x{c}} distance based centrality measures are the [{\em Shortest path\/}] \Closeness\ \cite{bavelas1948,bavelas1950} %freeman1978centrality
\beq{e:clos} %ref
f(u)=\Big(\sum_{v\in V}d(u,v)\Big)^{-1},\quad u\in V
\eeq
and [{\em Shortest path\/}] \Eccentricity\ \cite{bavelas1948,harary1953} %,HageHarary95
\beq{e:ecce} %ref
f(u)=(\max_{v\in V}d(u,v))^{-1},\quad u\in V.
\eeq
The heuristics behind \Closeness\ {is that the most central node should have the minimum sum of distances from all other nodes.} By adopting \Eccentricity, {we establish that the most central node minimizes the radius of the ``ball'' centered at that node and containing all other nodes.}

General classes of \Closeness\ and \Eccentricity\ \emph{\hl{centralities}\/} are defined by \eqref{e:clos} and \eqref{e:ecce} with $d(u,v)$ being {\em arbitrary\/} distances for graph nodes.
In the literature, several classes of such distances and, more generally, dissimilarity measures have been proposed (see, e.g., \cite{Che13Paris,avrachenkov2019similarities}). Substituting them\x{ for $d(u,v)$} in \eqref{e:clos} and \eqref{e:ecce} provides centralities with varying properties.\x{whose properties may vary (differ in part from those of the \Closeness\ and \Eccentricity\ induced by the shortest path distance).}
%and proximity measures \cite{CheSha98}.
Many of the alternative distances and dissimilarity measures are defined via \emph{graph kernels}.
\x{We now}Let us consider a few of them.

1. The parametric\x{ family of} {\it Katz\/} \cite{Katz53} {\it  kernels\/} (also referred to as {\it Walk\x{ proximities}\/}~\cite{CheSha98} or {\it Neumann diffusion\/} \cite{FoussSaerensShimbo16} {\it kernels}) are defined as
\beq{e:PWalk}
P^\Katz(t) = \sum_{k=0}^\infty (tA)^k = (I-tA)^{-1}
\eeq
with $0<t<(\rho(A))^{-1}.$ %, where $A=(a_{ij})$ is the adjacency matrix of $G$ and $\rho(A)$ is the spectral radius of~$A.$ %Neumann

2. The {\it Communicability kernels\/} \cite{estrada2005subgraph,FoussPirotte06Ker} are
$$%\beq{e:PComm}
P^\Comm(t) = \sum_{k=0}^\infty\frac{(tA)^k}{k!} = \exp(tA),\quad t>0.
$$%\eeq
%It is an instance of exponential diffusion kernels~\cite{KL02}.
%Since $K^\Comm$ is positive semidefinite, by Schoenberg's theorem, it can be transformed by \eqref{e_D2K} into a matrix of squared Euclidean distances.

Two other\x{ families} classes of kernels are defined similarly through the Laplacian matrix~\eqref{e:L}. %$L=L(G).$

3. The {\it Forest kernels}, or {\it regularized Laplacian kernels\/} \cite{CheSha95b,SmolaKondor03} are
\eq{e:PFor}{
P^{\For}(t) = (I + tL)^{-1},\;\; \mbox{where}\;\; t>0.
}

4. The {\it Heat kernels\/} are the Laplacian exponential diffusion kernels~\cite{KondorLafferty02diffusion}
$$%\beq{e:PHeat}
P^{\Hea}(t) = \sum_{k=0}^\infty\frac{(-tL)^k}{k!} = \exp(-tL),\quad t>0.
$$%\eeq

By Schoenberg's theorem \cite{Schoenberg35,Schoenberg38}, if matrix $P=(p_{uv})$ is a kernel (i.e., is positive semidefinite), then it  produces a\x{generates a squared} Euclidean distance %for graph nodes
$d(u,v)$ by means of the transformation
\beq{e:K2D2} %ref
d(u,v)=\sqrt{\tfrac12(p_{uu}+p_{vv}-p_{uv}-p_{vu})},\quad u,v\in V,  %\big)^{\frac12}
\eeq
where $\frac12$ is the scaling factor. Thereby $G$\/ has an exact representation consistent with $P$ in Euclidean space. A coordinate form of this representation can be found using multidimensional scaling. %This induces %obtained

Thus, all Walk, Communicability, Forest, and Heat kernels with appropriate parameter values give rise to distances\x{ $d(u,v)$} whose substitution into \eqref{e:clos} and \eqref{e:ecce} yields\x{ [{\em generalized\/}]} \Closeness\ and \Eccentricity\ centralities. We denote them by \Closeness(\xz{\em Kernel}\/) and \Eccentricity(\xz{\em Kernel}\/)\x{, respectively,} with the corresponding kernels substituted.

Furthermore, if $P_{n\times n}=(p_{uv})$ determines a {\it proximity measure\/} (\x{viz.,}which means that for any
%satisfies the triangle inequality for proximities
%$u,v,w\in \{1\cdc n\},$\; $p_{uv}+p_{uw}-p_{vw}\le p_{uu}$, and if $w=v$ and $v\ne u$, then the inequality
$x,y,z\in V\x{ \{1\cdc n\}},$\; $p_{xy}+p_{xz}-p_{yz}\le p_{xx}$, and this inequality
is strict whenever $z=y$ and $y\ne x$), then \cite{CheSha98a} transformation\x{ \eqref{e:K2D2}}
\eq{e:K2D}{d(u,v)=\tfrac12(p_{uu}+p_{vv}-p_{uv}-p_{vu}),\quad u,v\in V}
provides\x{generates} a distance function that satisfies the axioms of a metric. The Forest kernel with any $t>0$ produces a proximity measure, while kernels in the remaining three families do so when $t$ is sufficiently small~\cite{avrachenkov2019similarities}. The centralities obtained from a specific {\em Proximity\/} measure\x{ through} by transformation \eqref{e:K2D} and substitution of the resulting distance into \eqref{e:clos} and \eqref{e:ecce} will be denoted by \Closeness$^*($\xz{\em Proximity}\/$)$ and \Eccentricity$^*($\xz{\em Proximity}\/$)$, respectively.

Moreover, if $P$ represents a strictly positive {\it transitional measure on~$G$\/} (i.e., $p_{xy}\,p_{yz}\le p_{xz}\,p_{yy}$ for all $x,y,z\in V(G)$ with equality\x{and} $p_{xy}\,p_{yz} = p_{xz}\,p_{yy}$ whenever\x{iff} every path in $G$ from $x$ to $z$ visits $y$), then transformation
$$%\beq{e:ln}
\hat{p}_{uv}=\ln p_{uv},\quad u,v\in V
$$%\eeq
 %\eqref{e:ln}
produces \cite{Che11AAM,Che13Paris} a proximity measure. In this case, \eqref{e:K2D} applied to $\hat{P}=(\hat{p}_{uv})$ reduces to
\eq{e:logdist}{d(u,v)=\tfrac12(\ln p_{uu}+\ln p_{vv}-\ln p_{uv}-\ln p_{vu})}
and generates \cite{Che13Paris} a {\it cutpoint additive distance\/} $d(u,v),$ viz., such a distance that $d(u,v) + d(v,w) = d(u,w)$ whenever\x{iff} $v$ is a cutpoint between $u$ and $w$ in $G$ (or, equivalently, whenever\x{iff} all paths from\x{connecting} $u$ to $w$ visit~$v$).
The centralities obtained from any strictly positive {\!\it Transitional\!\_Measure\/}\x{ $P=(p_{uv})$} by substituting distance \eqref{e:logdist} into \eqref{e:clos} and \eqref{e:ecce} will be denoted by \Closeness$^*(${$\log$\it Transitional\!\_Measure}$)$ and \Eccentricity$^*(${$\log$\it Transitional\!\_Measure}$)$, respectively.

Since the Walk and Forest kernels determine \cite{Che11AAM} strictly positive transitional measures\x{ on the corresponding connected graph. Therefore}, \eqref{e:logdist} transforms them into cutpoint additive distances. Using\x{Substituting them into} \eqref{e:clos} and \eqref{e:ecce} we obtain\x{produces} \Closeness$^*(\log$\Forest$)$, \Closeness$^*(\log$\Walk$)$ and the corresponding \Eccentricity$^*(\cdot)$ centrality measures.

Thus, the above results allow us to define\x{{\it generalized}} \Closeness\x{-type} and\x{{\it generalized}} \Eccentricity\x{-type} centrality measures obtained by substituting, among others, the:
%\Up{.8}
\begin{itemize}%[noitemsep,topsep=0pt]
\item Forest kernel;
\item Heat kernel;
\item logarithmic Forest kernel;
\item logarithmic Walk kernel;
\item logarithmic Heat kernel, and
\item logarithmic Communicability kernel
\end{itemize}
%\Up{.5}
transformed by \eqref{e:K2D2} or \eqref{e:K2D} into \eqref{e:clos} and~\eqref{e:ecce}.
These centralities were used in a culling survey \cite{CheGub20How-} with parameter $t=1$ for the Forest, Heat, and Communicability kernels and $t=(\rho(A)+1)^{-1}$ for the Walk kernel.

Note that the authors of \cite{jin2019forest} examined the \Closeness$^*$(\xz\Forest) centrality and observed that the order of node importance induced by forest distances on some simple graphs was consistent\x{ is in agreement} with their intuition. Their conclusion was that ``forest distance centrality has a better discriminating power than alternate metrics such as betweenness, harmonic centrality, eigenvector centrality, and PageRank''.

In addition to the above approaches, kernels and proximity/similarity measures can yield centralities without converting to distances. An example is the \EstradaSubgraph\ centrality~\cite{estrada2005subgraph}. For a node $u,$ it is equal to the diagonal entry $p^\Comm_{uu}(t)$ of the Communicability kernel, so we denote it by \Communicability$(K_{ii}).$
Similarly, \Walk$(K_{ii})$ is the measure $\x{f(u)=}p^\Katz_{uu}(t),$ $u\in V$ determined by the diagonal entries of the Walk kernel.

An alternative is to sum the off-diagonal row elements of a kernel matrix. For example, \Communicability($K_{ij}$) and \Walk($K_{ij})$ are the centralities defined by $f^t(u)=\sum_{v\ne u}p^\Comm_{uv}(t)$ and $f^t(u)=\sum_{v\ne u}p^\Katz_{uv}(t),$ $u\in V,$ %\Rem{[$v\ne u,$ really? $t=$?]}
respectively.

Finally, \TotalCommunicability\ \cite{benzi2013total} is calculated by summing {\em all\/} row entries of the Communicability kernel:
$f^t(u)=\sum_{v\in V}p^\Comm_{uv}(t)=(P^\Comm(t)\xy\bm1)_u$; it can be described \cite{deMeoProvetti2019general} in terms of ``potential gain''.
\x{as well as }The \x{ corresponding \Katzc\ }\TotalWalk\ measure $f^t(u)=\sum_{v\in V_{\mathstrut}}p^\Katz_{uv}(t)=(P^\Katz(t)\xy\bm1)_u$ is order equivalent to the famous Katz centrality \cite{Katz53} $f^t(u)=((P^\Katz(t)-I)\bm1)_u$, which we consider in Section~\ref{s:Self-c}.
\begin{figure}[!t] %[!ht]
\begin{center}
\includegraphics[width=1.01\linewidth]{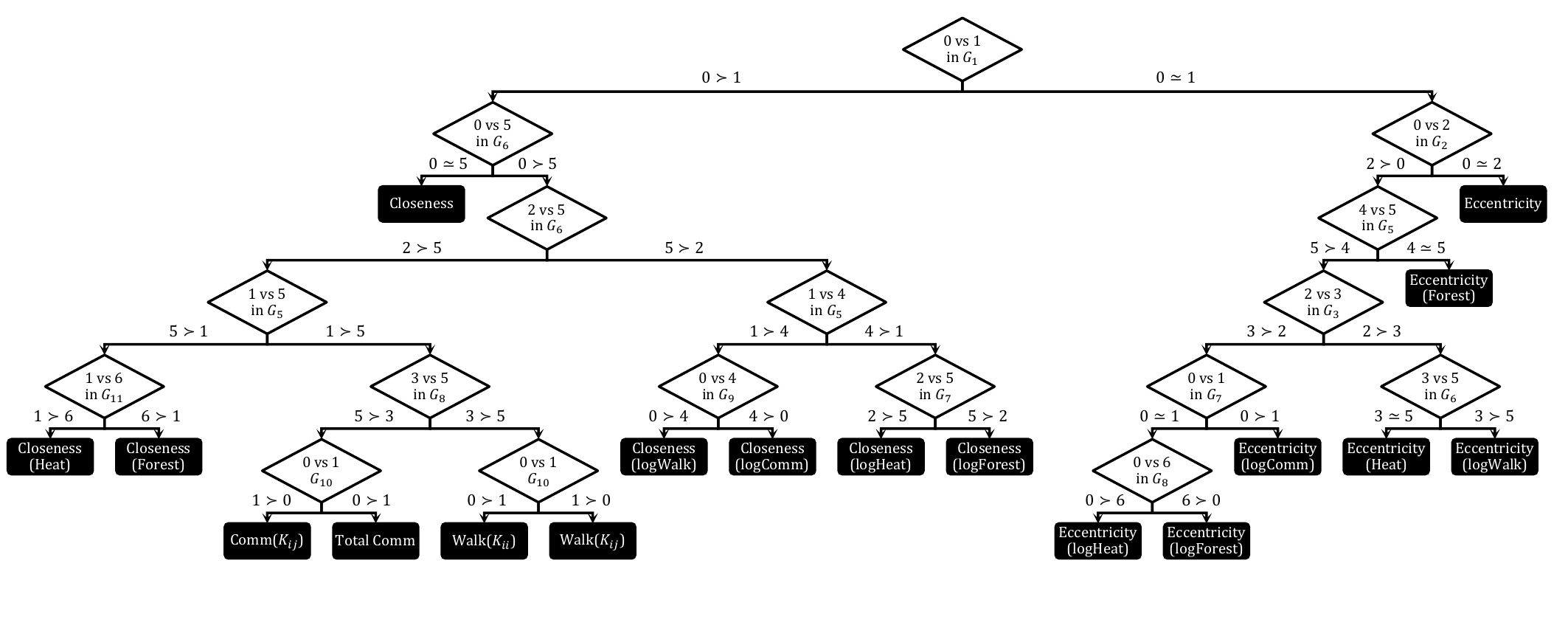}
\includegraphics[width=.83\linewidth]{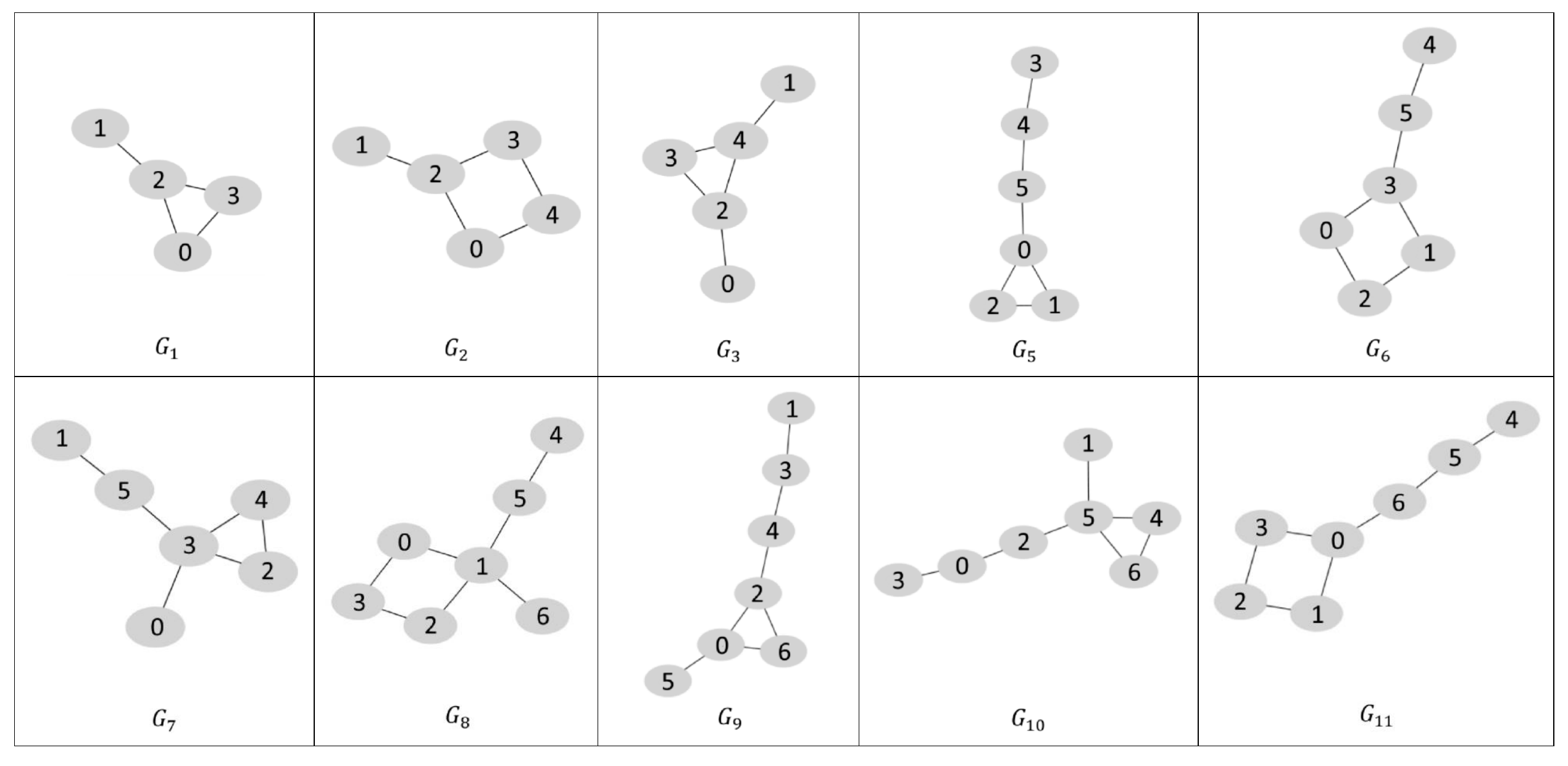}
\caption{Order differences between 16 kernel-based centralities, {\em Shortest path closeness}, and {\em Shortest path eccentricity}. In this figure, $i\simeq j$ means that nodes $i$ and $j$ have the same centrality, $i\succ j$ that $i$ has a higher centrality than $j$; ``Comm'' is short for Communicability.
\aB{The dendrogram of measures is constructed as described in \cite{CheGub20How-}} \hl{with the graph numbers preserved}. \label{f:Mtree}}
\end{center}
\end{figure}

Let us compare 16 kernel-based measures with the parameters\x{ specified} mentioned above and the classical \Closeness\ and \Eccentricity\ centralities by examining how they act on ten simple graphs generated \hl{using the procedure described in} \cite{CheGub20How-} (Fig.~\ref{f:Mtree}).
The \Closeness, $K_{ii}$, $K_{ij},$ and {\em Total\/} measures give node $0$ in $G_1$ a higher centrality than node $1$, as shown in Fig.~\ref{f:Mtree} by $0\succ1$.
Each \Eccentricity\ measure assigns the same (minimal) centrality to all the diameter ends\footnote{If $d:V\times V\to\R_+$ is a distance (or more generally a dissimilarity measure) for the nodes of $G$, then $\max_{u,v\in V}d(u,v)$ is referred to as the corresponding {\em diameter\/} of $G$, while the nodes $u$ and $v$ that maximize $d(u,v)$ are the {\em diameter ends}.} of a graph. As a result, the \Eccentricity\ measures assign the same centrality to nodes 0 and 1 of $G_1$, which is shown as $0\simeq1$. The {\em Shortest path eccentricity}\x{ assigns} sets $0\simeq2$ in $G_2$, whereas the other {\em Eccentricities} set $2\succ0$. \Eccentricity(\Forest) with $t=1$ assigns the highest centrality to both $4$ and $5$ in $G_5$, the others\x{ only to $5$} set $5\succ4$.
Turning to \Closeness\ measures, a singularity\x{peculiarity} of the {\em Shortest path closeness\/} is that it equalizes the centrality of $0$ and $5$ in $G_6$, while its competitors set $0\succ5$.
Furthermore, all logarithmic \Closeness\ measures in this test set suggest $5\succ2$ in $G_6$, while the other measures establish $2\succ5.$ {By looking at the separating graphs, the reader can judge which measures offer the most reasonable rankings.}
%the Eccentricities differently rank $3$ and $2$ in $G_3$.
%Fig.~\ref{f:Mtree} demonstrates how the graphs $G_1$ to $G_3$ and  $G_5$ to $G_{11}$ separate these measures. Consider two****** examples of separation.

\section{Axioms of Bridge and Self-consistency}
\label{s:axioms}

The axioms in this section determine the relation between the centrality values of a pair of\x{ two} nodes in a\x{ fixed} graph of a special structure.
%The axioms considered in this section apply to centrality measures $f$ acting on a fixed graph~$G.$
%
As mentioned above, the measures under study assign centrality to nodes based solely on the graph structure. %~\cite{RoyTredan10SharpCentr}
The\x{ following} \Equivalence\ axiom is a\x{ simple particular} partial embodiment of this idea (cf.\ \cite[\Axiom~A3]{sabidussi1966centrality}).

\begin{axiom}[\Ee: \Equivalence]{ If $u,v\in V(G)$ and $u\sim v,$ then $f(u)=f(v).$}\end{axiom}

All measures under consideration satisfy \Axiom\footnote{We use the notation \Ee\ so as not to confuse this axiom with the edge set $E$ of a graph.}~\Ee; it will be assumed by default.
%}

Among the most appealing axioms characterizing\x{ restricting} various classes of reasonable centrality measures are those of an ordinal nature.
Such axioms\x{ make it possible} allow one to compare the centrality of some nodes, but they do not\x{ determine specific computational algorithms (for calculating centrality in the general case)} indicate how to calculate the centrality values. In this sense\x{other words}, they\x{ do not look like} are not fingerprints of\x{ specific} particular centrality indices.

%In many axiomatic constructions, conditions of Positive responsiveness\x{ or Monotonicity} are of primary importance.
{\em Positive responsiveness\/} is a type of axiom, which is of primary importance in many axiomatic constructions.
The\x{ common (generic) framework} template of these axioms is: ``an increase in input (making a node more central from some point of view) leads to\x{ yields (causes)} an increase in output (i.e., raises its centrality value or rank)''.
Now we present two axioms of this kind. In the next two sections, we will find centrality\x{ indicate several} measures that satisfy them.\x{ demonstrate how they can help replace the complete survey with a short one.} % a fairly long

Recall that a {\em bridge\/} in a graph is an edge whose deletion increases the number of graph's connected components.\x{ and that we consider connected graphs.}
The following axiom \cite{skibski2018axioms} relates the centrality of the end-nodes\x{C} of any bridge. %We consider a stronger version of this axiom (No!).

\begin{axiom}[B: Bridge]{ If $G$ is connected and the removal of edge $\{u,v\}$ from $E(G)$ separates\x{ splits}
$G$ into two connected components with node sets denoted by $V_u\ni u$ and $V_v\ni v$ (i.e., $\{u,v\}$ is a bridge\x{ in $G$}), then $|V_u|<|V_v|\ToTo$ $f(u)<f(v).$} %if and only if
\end{axiom}

A strengthening of this axiom is the {\em Ratio property\/}~\cite{Khmelnitskaya23}, which holds when for a positive $f,$ under the same premise,\x{ $f(w)>0$ for all $w\in V$ and} $f(u)/f(v)=|V_u|/|V_v|.$
\medskip

The idea behind the second axiom is quite different.
One may assume\x{ believe} that the vector of centrality values of the \emph{neighbors\/} of any\x{ arbitrary} node $u$ carries a lot of information about the centrality of $u$ itself (cf.\ Consistency in~\cite{dequiedt2017local}).
A more specific form of this idea is that ``\xz\emph{the higher the centrality values of a node's neighbors$,$ the higher the centrality of the node itself}''.

This is in line with the justification given by Bonacich and Lloyd \cite{BonacichLloyd01} to the \Eigenvector\ centrality, a measure satisfying (see Section~\ref{s:Self-c} below) the axiom we are approaching now:
``The eigenvector is an appropriate measure when one believes that actors' status is determined by those with whom they are in contact. This conception of importance or centrality makes sense in a variety of circumstances. Social status rubs off on one's associates. Receiving information from knowledgeable sources adds more to one's own knowledge. However, eigenvectors can give weird and misleading results when misapplied.''
%\medskip

This idea is exactly embodied in the Self-consistency axiom.
For directed graphs, it was used in \cite{CheSha97a,CheSha98AOR,CheSha99,DiazHendrickxLohmann13SCW,Csato19ImposPC,Csato19ImposT,Csato19Journal}, in
the case of undirected graphs, in \cite{bandyopadhyay2018generic}.

\begin{axiom}[S: Self-consistency]{ If nodes $u,v\in V$ have the same degree and there exists a bijection between $N_u$ and $N_v$ such that each\x{every} element of $N_u$ is, according to $f,$ no more central than the corresponding element of $N_v$, then $f(u)\le f(v).$ %$u$ is no more central than~$v.$
If ``no more'' is actually ``less'' at least once, then $f(u)<f(v).$}
\end{axiom}

Both the Bridge and Self-consistency axioms belong to the class of positive responsiveness axioms, however,\x{ in contrast to the Bridge axiom,} the positivity requirement in the premise of Self-consistency is not objective: it is\x{ reduces to} positivity in terms of $f$ itself. \x{As a result,}This implies that whenever $f$ satisfies \Axiom~S and the values of $\bar f$ are ordered oppositely to those of $f$\x{ (for example, $\bar f=1/f$, where $f>0$)}, it holds\x{we have} that $\bar f$ also satisfies~S. Thus, Self-consistency equally welcomes both centrality and peripherality measures. Consequently, while the \emph{sole\/} \Axiom~S allows in some cases to conclude that $f(u)=f(v),$ it never implies~$f(u)>f(v).$ In particular, a centrality that states $f(u)=f(v)$ for all $u,v\in V,$ trivially\x{obviously} satisfies \Axiom~S for any graph. That is why Self-consistency is usually combined with other axioms that specify centrality enhancers in terms of graph structure, not just reflections of neighbors' centrality\x{ in accordance with {\em Self-}consistency}. These axioms include the edge-monotonicity conditions discussed in Section~\ref{s:Monot}.

In the following two sections, we present several results on the centrality measures that satisfy the Bridge or Self-consistency axiom.

\section{Centrality measures satisfying the Bridge axiom}
\label{s:Bridge}

In the statements of this section, we use the notion of cutpoint additive distance and the \Closeness$^*(\log$\Forest$)$ and \Closeness$^*(\log$\Walk$)$\x{ centrality} measures introduced in Section~\ref{s_NewMes}.

\begin{lemma}\label{l:CutBridge}
Any\x{ generalized} \Closeness\ centrality of the form \eqref{e:clos} such that the corresponding distance $d(\cdot,\cdot)$ is cutpoint additive satisfies\x{ the Bridge} \Axiom~B.
\end{lemma}

\begin{proof}
For any connected $G,$ let $f(u)=\left(\sum_{v\in V}d(u,v)\right)^{-1}$ be a \Closeness\ centrality such that $d(\cdot,\cdot)$ is a cutpoint additive distance. Suppose that $\{u,v\}$ is a bridge in~$G.$ Since $v$ is a cutpoint between $u$ and any node $w\in V_v\sssm\{v\},$ it holds that
\eqss{
\nonumber
(f(u))^{-1}
&=&\sum_{w\in V(G)}d(u,w)
 =\sum_{w\in V_u}d(u,w)
 +\sum_{w\in V_v}d(u,w)\\\nonumber
&=&\sum_{w\in V_u}d(u,w)+|V_v|\,d(u,v)
+\sum_{w\in V_v}d(v,w).
}
Similarly,
$(f(v))^{-1}
%=\sum_{w\in V(G)}d(v,w)
=\sum_{w\in V_v}d(v,w)+|V_u|\,d(v,u)+\sum_{w\in V_u}d(u,w).$
\x{Therefore Consequently,}Hence
$$
(f(u))^{-1}-(f(v))^{-1}%=\sum_{w\in V(G)}d(u,w)-\sum_{w\in V(G)}d(v,w)
=(|V_v|-|V_u|)\,d(u,v),
$$
consequently, $f(u)<f(v)\x{ \ToTo (f(v))^{-1}<(f(u))^{-1}} \ToTo |V_u|<|V_v|.$ Thus, $f$ satisfies the Bridge axiom.
%\qed
\end{proof}

\medskip
The \Connectivity\ centrality \cite{Khmelnitskaya23} of vertex $u$ is equal to the number of permutations $\pi=(\pi_1\cdc\pi_{|V|})$ of $V(G)$ such that $(i)$~$\pi_1=u$ and $(ii)$~for each\x{every} $j\in\{2\cdc|V|-1\},$ the induced subgraph of $G$ with node set $\{\pi_1\cdc\pi_j\}$ is connected.

\begin{proposition}
The Shortest path \Closeness$,$ \Connectivity$,$ \Closeness$^*(\log$\Walk$),$ and \Closeness$^*(\log$\Forest$)$\x{ centralities} satisfy the Bridge axiom.
\end{proposition}

\begin{proof}
The fulfilment of the Bridge axiom for the Shortest path \Closeness\/\x{ was proved in}\ is due to Skibski and Sosnowska~\cite{skibski2018axioms}.
Alternatively, it follows from Lemma~\ref{l:CutBridge}.

The Bridge axiom holds for \Connectivity\ since this centrality measure satisfies \cite{Khmelnitskaya23} the\x{ stronger} Ratio property mentioned in Section~\ref{s:axioms}.

The Walk \eqref{e:PWalk} and Forest \eqref{e:PFor} kernels represent \cite{Che11AAM} strictly positive transitional measures on any connected graph. Therefore,\x{ definition} \eqref{e:logdist}\x{ {e:K2D}} transforms \cite{Che13Paris} them\x{ corresponding logarithmic kernels} into cutpoint additive distances $d(u,v).$
By Lemma~\ref{l:CutBridge} this implies that the\x{ generalized} \Closeness\ centralities based on\x{ corresponding to} these distances, namely, the  \Closeness$^*(\log$\Walk$)$ and \Closeness$^*(\log$\Forest$)$ centralities, satisfy the Bridge axiom.
\end{proof}

\medskip
Similarly, other strictly positive transitional measures \cite{Che11AAM} and cutpoint additive distances also produce %enable one to construct
centralities that satisfy the Bridge axiom.
\begin{figure}[!t] %[!ht]
\begin{center}
\includegraphics[height=11em]{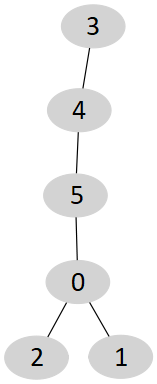}
\caption{A tree on which \Betweenness\ centrality violates \Axiom~B.\label{f:Btree}}
\end{center}
\end{figure}

\begin{remark}\label{r:BetwBri}
It is worth noting that the \Betweenness\ centrality~\cite{freeman1977set} satisfies the Bridge axiom for many graphs, however, this is not the case in general.
A simple (and, as is easy to prove, minimal) graph on which \Betweenness\ violates this axiom is shown in Fig.~\ref{f:Btree}.
Here, \Axiom~B requires that the centrality of nodes $0$ and $5$ be equal since $|V_0|=|V_5|$. However, the \Betweenness\ centrality of node $0$ is higher than that of node $5$, since $0$ lies on the shortest path from $1$ to~$2.$
\end{remark}

\Up{.5}
\begin{remark}\label{r:Kern}
{The Bridge axiom forces us to ignore the difference in edges on subsets $V_u$ and $V_v$ of the same cardinality. Therefore, centrality measures that rank nodes\x{ taking} based on this difference\x{ into account} violate} \Axiom~B (see Table~1 in \cite{CheGub20How-} and Remark~\ref{r:BnoT}).
\end{remark}

\section{Centrality measures satisfying Self-consistency}
\label{s:Self-c}

\x{In this section,}
%\medskip
To formulate a necessary and sufficient condition for Self-consistency, we introduce two definitions.

\begin{definition}\label{d:ScorFun}
A\x{ real-valued} function $\ph\!:\M_k\to\R,$ where $\M_k=\{M\!:\,0<|M|\le k\},$ $M$ being a multiset%
\footnote{A finite nonempty multiset of elements of a set $\mathcal X$ is an equivalence class of tuples with components from $\mathcal X$ such that two tuples $\bm z$ and $\bm z'$ are equivalent whenever $\bm z'$ can be obtained from $\bm z$ by permuting its components. As distinct from a set, a multiset may contain several copies of the same element, because some components of a tuple may coincide.}
of real numbers, will be called a {\em scoring function\/} if $\ph(M)$ is strictly increasing in each element of $M,$ while the remaining elements\x{ of $M$}, including those equal to the varying one, are fixed.\x{provided that all other arguments are fixed.}
\end{definition}

In the following definition, a {\em centrality vector\/} associated by a centrality measure $f$\/ with a graph $G$ having $V(G)=\{1\cdc n\}$ is a vector $\bm x=(x_1\cdc x_n)^T$ such that $x_u=f(u),$ $u\in V(G)\x{=1\cdc n}.$

\begin{definition}\label{d:NeighMon}
A centrality vector $\bm x=(x_1\cdc x_n)^T$ associated with a connected graph $G$ such that $V(G)=\{1\cdc n\}$
%($x_u=f(u),$ $u\in V(G),$ where $f$ is the corresponding centrality measure)
{\em has a supporting neighborhood representation\/} if there exists a scoring function $\ph$ such that $\bm x$ satisfies the system of equations
\eq{e:deMon}{
x_u=\ph(\{x_w\!:\,w\in N_u\}), \quad u=1\cdc n.
}
\end{definition}

In Definition~\ref{d:NeighMon}, $\{x_w\!:\,w\in N_u\}$ is the multiset of the components of $\bm x$ that correspond to the neighbors of node $u$ in~$G.$ If a centrality vector has \amonotonic\ neighborhood representation, then the centrality of each node is uniquely (and increasingly\x{c}) determined by the centralities of its neighbors.\x{ and so this vector it is a solution to\x{ finding this vector reduces to solving} the system~\eqref{e:deMon}.} A related concept of monotone implicit form appeared earlier in~\cite{CheSha98AOR} and was used in \cite[Theorem~12]{CheSha99}.

\begin{lemma}\label{l:monSC}
A centrality measure on $G$ satisfies Self-consistency if and only if the centrality vector this measure associates with $G$ has \amonotonic\ neighborhood representation.
\end{lemma}

\begin{proof}
Suppose that the centrality vector $\bm x=(x_1\cdc x_n)^T$ associated with\x{ a connected graph} $G$ has \amonotonic\ neighborhood representation\x{ of the form}~\eqref{e:deMon}. Let the premise of Self-consistency be true for nodes $u$ and~$v.$ Consider the equations \eqref{e:deMon} corresponding to $u$ and~$v$\xy:
\eqss{\label{ee:deMon1}
x_u&=&\ph(\{x_w:\,w\in N_u\}),\\\label{ee:deMon2}
x_v&=&\ph(\{x_w:\,w\in N_v\}).
}

Since there is a bijection that maps each element of $N_u$ to an element of $N_v$ with a greater or equal centrality,\x{ sequentially} step by step replacing in \eqref{ee:deMon1} the\x{ centrality} $x_w$ value of each element of $N_u$ by the\x{ centrality value} $\bm x$ component of the corresponding element of $N_v$ and using the definition of \monotonic\ neighborhood representation, we get a growth or preservation of the value of $\phcc$ at each step, yielding the value $x_v$ in the last step. This implies that $x_u\le x_v,$ or, stronger, $x_u<x_v$ provided that\x{ whenever} $x_w$ has been strictly increased at least once.\x{ of these transformations differs from the identity.} Therefore, Self-consistency is satisfied.

Conversely, suppose that a centrality measure on $G$ is Self-consistent.
Let us construct\x{ prove that there\x{ exists} is} a scoring function $\phc$ that provides \amonotonic\ neighborhood representation of the centrality vector $\bm x=(x_1\cdc x_n)^T$ associated with~$G.$ First, we set $\ph(\{x_w\!:\,w\in N_u\})\stackrel{\mathrm{def}}{=}x_u$ for all $u\in\{1\cdc n\}.$
%This\x{ does not contradict (obeys)} fits the definition of a scoring function due to Self-consistency. In particular,
Whenever $\{x_w\!:\,w\in N_u\}=\{x_w\!:\,w\in N_v\}$ for some $u,v\in V,$ Self-consistency implies $x_u=x_v,$ i.e., our\x{ the above} definition of $\phc$
on the set of multisets $\PP=\{\{x_w:\,w\in N_u\},$ $1\le u\le n\}\subset\M_k$
%on the arguments of the form $\{x_w:\,w\in N_u\},$
is not contradictory\x{consistent}.
Thus, we defined a function $\phPc=\phc$ on $\PP,$ and to obtain \amonotonic\ neighborhood representation of $\bm x,$ it suffices to extend  $\phPc$ to the entire set $\M_k$ $\left(k=\max\{|N_u|,1\le u\le n\}\right)$ of multisets\x{$M$ of non-negative} of real numbers in such a way that the resulting $\phc$ is strictly increasing on~$\M_k.$

By the definition of a scoring function, the strict increase of $\phc$ is required with respect to the following\x{ strict partial} preorder $\succE$ on $\M_k$: for $\xx,\yy\in\M_k,$ $\xx\succE\yy\ToTo$ \x{$[\xx\ne \yy$ and }[there is a bijection between $\xx$ and $\yy$ such that every element of $\yy$ does not exceed the corresponding element of~$\xx].$ The condition of strict increase reduces to the\x{ conjunction of two} implication $[\xx\succE\yy \mbox{ and }\yy\not\succE\xx]\To\ph(\xx)>\ph(\yy),$ since the second necessary implication $[\xx\succE\yy \mbox{ and }\yy\succE\xx]\To\ph(\xx)=\ph(\yy)$ is trivial because\x{ as} its premise implies in our case that $\xx=\yy.$

Observe that the preorder $\succE$ has a numerical [utility] representation. This means that there exists a function $\uti\!:\M_k\to\R$ such that for all $\xx,\yy\in\M_k,$
$\xx\succ \yy\To \uti(\xx)>\uti(\yy),$
%{and} $\xx\approx \yy\To \uti(\xx)=\uti(\yy),$
where$,$ by definition$,$
$\xx\succ \yy \ToTo [\xx\succE\yy \mbox{ and }\yy\not\succE\xx].$
Indeed, $\uti(X)$ can be defined\x{, for example}, say, as the sum of the elements of multiset~$X,$ which guarantees $\xx\succ \yy\To \uti(\xx)>\uti(\yy)$\x{ by definition of~$\succ.$}
and so $\uti$ is a numerical representation of~$\succE$\xy.

By Self-consistency, $\phPc$ strictly increases on~$\PP$\x{ the set of multisets $\PP=\{\{x_w:\,w\in N_u\},$ $1\le u\le n\}\subset\M_k$}, i.e.,
\x{This means that $\phc=$}$\phPc$ is a numerical representation of $\succE_\PP,$ the restriction of $\succE$ to~$\PP.$
Since $\succE$ has a numerical representation, it follows from \cite[Theorem~1]{Che22Util-} that $\phPc$ has a strictly increasing extension to $\M_k$ if and only if $\phPc$ is gap-safe increasing, i.e., is weakly increasing and for any $\xx,\yy\in\M_k\cup\{\mainfty,\painfty\},$ $\yy\succ\xx$ implies
\eq{e:SepMo}{%\nonumber
\inf\xy\{\phP(\zz)\xz:\,\zz\succE\yy,\,\zz\in\PP\}\,>\,
\sup\xy\{\phP(\zz)\xz:\,\xx\succE\zz,\,\zz\in\PP\},}
where, %$\zz\succE\yy$ means $[\zz\succ\yy$ or $\zz=\yy]$
by definition, $\mainfty$ and $\painfty$ are elements that satisfy $\painfty\succ \zz\succ\mainfty$ for all $\zz\in\M_k$
and, by convention, $\sup\varnothing=-\infty$ and $\inf\varnothing=+\infty$ (cf.~\cite[Section~4]{Tanino88supremum}).

To prove that $\phPc$ is gap-safe increasing, first observe that since $\PP$ is finite, $\sup$ and $\inf$ in \eqref{e:SepMo} can be replaced by $\max$ and $\min,$ respectively,\x{ provided that we set} under the convention that $\max\varnothing=-\infty$ and $\min\varnothing=+\infty.$ Then, if the [multi]sets on the left-hand and right-hand sides of \eqref{e:SepMo} are both nonempty, then for any $\zz''$ and $\zz'$ minimizing $\phP(\zz)$ on the left and maximizing $\phP(\zz)$ on the right, respectively, $\zz''\succE\yy\succ\xx\succE\zz'$ holds, and by the mixed strict transitivity\footnote{This means that for any $\xx,\yy,\zz\in\M_k,\xy$ $\zz\succE\yy\succ\xx\To\zz\succ\xx\xy$ and $\xy\yy\succ\xx\succE\zz\To\yy\succ\zz.$}
of $\succE,$ $\zz''\succ\zz'.$ By Self-consistency this implies $\phP(\zz'')>\phP(\zz')$ and so\x{c} \eqref{e:SepMo} is true\x{is valid}. Otherwise, if some multiset in \eqref{e:SepMo} is empty, then we have $+\infty$ on the left or/and $-\infty$ on the right, in a possible combination with a finite number on the other side\x{ one of the sides}. In all these cases, \eqref{e:SepMo}\x{ holds true} is valid, hence $\phPc$ is gap-safe increasing. Thus, by \cite[Theorem~1]{Che22Util-}, $\phPc$ can be extended to $\M_k$ so that its extension $\phc$ is a strictly increasing function and therefore, provides \amonotonic\ neighborhood representation of the centrality vector $\bm x=(x_1\cdc x_n)^T.$ This completes the proof. \x{Note that }The extension of $\phPc$ to $\M_k$ can be defined\x{ made}, in particular, using the approach proposed\x{ developed} in~\cite{Che22Util-}.
%\qed
\end{proof}

%\medskip
\smallskip
Lemma~\ref{l:monSC} will be used to prove the following statements, which involve five centrality measures; we now recall their definitions using the notation introduced in Section~\ref{s:notat}.

\smallskip
For a connected graph $G$ of order $n,$ %with spectral radius $\rho(A)$ of the adjacency matrix $A,$
vector $\bm x=(x_1\cdc x_n)^T$ presents:\x{ the}
\begin{itemize}
\item
the \KatzcCentrality~\cite{Katz53} if
\eq{e:K}{
\bm x=\sum_{k=1}^\infty(tA)^k\xy\bm1 = ((I-tA)^{-1}-I)\xy\bm1,
}
where $t\in\R$ is a parameter such that $0<t<(\rho(A))^{-1};$ %$\rho(A)$ being the spectral radius of $A,$
\item
the \BonacichCentrality~\cite{Bonacich87} with real parameters $\a$ and $\be>0$ if $\bm x$ satisfies the system of equations
\eq{e:B1}{
x_u=\sum_{w\in N_u}(\a+\be x_w),\quad u=1\cdc n;
}
\item
%For any graph $G,$
the \GeneralizedDegreeCentrality\xz~\cite{csato2017measuring} if $\bm x$ satisfies the system of equations
\eq{e:Ge_d}{
(I+\ve L)\bm x=\bm d,
}
where %$\bm d=(d_1\cdc d_n)^T=A\bm 1$ is the vector of node degrees and
$\ve>0$ is a real parameter;
\item
the \EigenvectorCentrality~\cite{landau1895relativen,bonacich1972factoring} if $\bm x$ is positive and satisfies the equation
\eq{e:eigen}{
A\bm x=\rho(A)\xy\bm x; %where $\rho(A)$ is the spectral radius of~$A;$
}
\item
the \PageRankCentrality~\cite{BrinPage1998anatomy} if $\bm x$ is positive and satisfies the equation
\eq{e:PR}{
\bm x=   \left(\a A^T(\diag(A\bm1))^{-1}+(1-\a)J\right)\xz\bm x,
}
where $J=\frac1n\bm{11}^T,$ while $\a\in\R$ is the {\em damping factor} (or {\em teleportation parameter}) satisfying\x{such that} ${0<\a<1}.$
{In the case of undirected graphs considered in this paper, $A^T=A.$}
\end{itemize}

\smallskip
It should be noted that substituting $\a=1$ in \eqref{e:PR} yields the {\em Seeley centrality\/} \cite{Seeley49Soc} whose scores are proportional \hl{to the node degrees} in the undirected case.
\GeneralizedDegreeCentrality\ results by applying the Generalized row sum method \cite{Che94}, whose score vector is proportional to $\bm x=(I+\a L)^{-1}\bm s,$ where $\bm s$ is the vector of row sums in a matrix of paired comparisons, to the adjacency matrix $A$ with row sum vector~$\bm d.$
\BonacichCentrality\ is a\x{ variant} version of \KatzcCentrality, therefore the name {Katz-Bonacich centrality\/} can often be found.

\begin{proposition}\label{p:SC}
The \GeneralizedDegree$,$ \Katzc$,$ \Eigenvector$,$ and \Bonacich\ centralities satisfy Self-consistency.
\end{proposition}

\begin{proof}
1.
  \x{For any graph $G,$ the \GeneralizedDegree~\cite{csato2017measuring} centrality vector $\bm x=(x_1\cdc x_n)^T$ satisfies the system of equations
  $$%\beq{e:Ge_d}
  (I+\ve L)\bm x=\bm d,
  $$%\eeq
  where $\bm d=(d_1\cdc d_n)^T=A\bm 1$ is the vector of node degrees and $\ve>0$ is a parameter}
Since $d_u=|N_u|$ for each $u\in V,$ Eq.~\eqref{e:Ge_d} can be written in component form as
\beq{e:Ge_ds}\nonumber %-
x_u(1+\ve|N_u|)-\ve\sum_{w\in N_u}x_w=|N_u|,\quad u=1\cdc n,
%x_u=\sum_{w=1}^n a_{uw}(\ve x_w-\ve x_u+1),\quad u=1\cdc n.
\eeq
which is equivalent to
\beq{e:Ge_ds1} %
x_u=(1+\ve|N_u|)^{-1}\sum_{w\in N_u}(1+\ve x_w),\quad u=1\cdc n.
%x_u=\sum_{w=1}^n a_{uw}(\ve x_w-\ve x_u+1),\quad u=1\cdc n.
\eeq

Eq.\:\eqref{e:Ge_ds1} is \amonotonic\ neighborhood representation of vector $\bm x$, therefore, by Lemma~\ref{l:monSC}, the \GeneralizedDegree\ centrality satisfies Self-consistency.

2.\x{(For any graph $G,$) The \Katzc\ centrality vector $\bm x=(x_1\cdc x_n)^T$ is expressed as
$$%\beq{e:K}
\bm x=\sum_{k=1}^\infty(tA)^k\bm1 = ((I-tA)^{-1}-I)\bm1,
$$%\eeq
where $0<t<(\rho(A))^{-1},$ $\rho(A)$ being the spectral radius of $A,$}
It follows from \eqref{e:K} that
\beq{e:K1} %
(I-tA)\bm x=t\bm d,\nonumber
\eeq
%where $\bm y=(y_1\cdc y_n)^T=\bm x+\bm 1,$ and so
which yields
\beq{e:K2}
%y_u=1+t\sum_{w\in N_u}y_w,\quad u=1\cdc n.
x_u=t\sum_{w\in N_u}(1+x_w),\quad u=1\cdc n.
\eeq

Since for any $t>0,$ \eqref{e:K2} is \amonotonic\ neighborhood representation of $\bm x$, Lemma~\ref{l:monSC} implies that the \Katzc\ centrality satisfies Self-consistency.

3.\x{For a connected graph $G,$ the \Eigenvector\ centrality is defined by $A\bm x=\rho(A)\bm x$ or,}
A component form of \eqref{e:eigen} is
\beq{e:E1}
x_u=(\rho(A))^{-1}\sum_{w\in N_u}x_w,\quad u=1\cdc n,
\eeq
which is \amonotonic\ neighborhood representation of $\bm x$. Hence, by Lemma~\ref{l:monSC}, the \Eigenvector\ centrality satisfies Self-consistency.

4.\x{For any connected graph $G,$ the \Bonacich\ centrality with parameters $\a$ and $\be>0$ satisfies the system of equations
\beq{e:B1}
x_u=\sum_{w\in N_u}(\a+\be x_w),\quad u=1\cdc n,
\eeq}
Equations \eqref{e:B1} of \Bonacich\ centrality provide \amonotonic\ neighborhood representation of~$\bm x$. By Lemma~\ref{l:monSC}, these centralities satisfy S\x{Self-consistency}. Comparing\x{It follows from the comparison of} \eqref{e:K2} and \eqref{e:B1} we observe that the \Katzc\ centralities\x{ belong to the class of} are the \Bonacich\ centralities with $\a=\be=t.$
On the other hand, $\a$ just scales $\bm x,$ so every vector of \Bonacich\ centralities with $\be=t$ is proportional to the vector of \Katzc\ centralities with parameter~$t.$
\end{proof}

\begin{figure}[ht] %\x{ht} F5
	\centerline{\includegraphics[width=6.9cm]{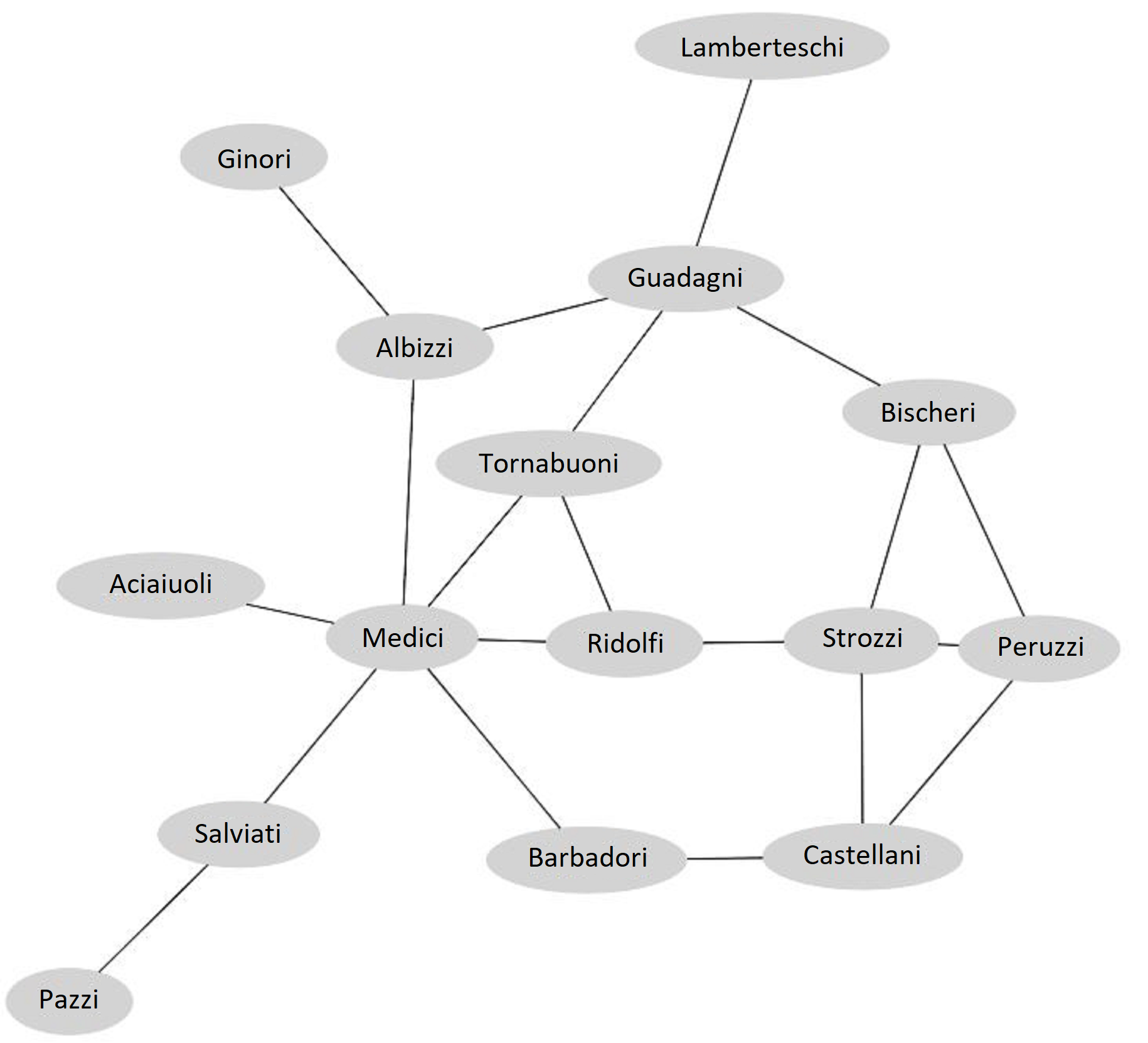}}   %[width=10cm]{Florence.png} %8.9cm
	\caption{Marriage network of the Florentine ruling families of the 15th {century} \cite{PadgettAnsell93} (without the isolated Pucci family).}
	\label{f:Floren}
\end{figure}

\smallskip
To prove that a centrality measure satisfies Self-consistency, it suffices to find its \monotonic\ neighborhood representation\x{ for it}, as we did, e.g., for the \Katzc\ centrality. Disproving this hypothesis\x{ of the Self-consistency of a measure} reduces to\x{ boils down to} giving a refuting example in the form of a\x{n appropriate} pair of nodes in\x{ a real-world or artificial} some graph\x{ network} that violates\x{ this axiom} Self-consistency. \x{Here, In some cases (among others)}For many measures, the well-known network of Florentine ruling families (Fig.~\ref{f:Floren}) can help with this, as we show in Lemma~\ref{l:5} and Proposition~\ref{p:9}.

Let $f$ be a centrality measure on a graph $G.$ For $k\in\N$ such that $1<k<n,$ we say that two $k$-tuples $(u_1\cdc u_k)$ and $(v_1\cdc v_k)$ each consisting of distinct nodes of $G$ are \emph{$f$ order equivalent\/} iff for any $i,j\in\{1\cdc k\},$ $\sign(f(u_i)-f(u_j))=\sign(f(v_i)-f(v_j)).$
\begin{lemma}\label{l:5} %5.
If a centrality measure $f$ satisfies \Axiom~S$,$ then for the Florentine families graph \cite{PadgettAnsell93} of Fig.~$\ref{f:Floren},$ the following tuples of nodes are $f$ order equivalent\/$:$\\
(a) $($Tornabuoni$,$ Albizzi$)$ and $($Ridolfi$,$ Ginori$);$\\
(b) $($Bischeri$,$ Peruzzi$)$ and $($Guadagni$,$ Castellani$);$\\
(c) $($Bischeri$,$ Castellani$)$ and $($Guadagni$,$ Barbadori$);$\\
(d) $($Peruzzi$,$ Castellani$)$ and $($Bischeri$,$ Barbadori$);$\\
(e) $($Tornabuoni$,$ Ridolfi$)$ and $($Guadagni$,$ Strozzi$);$\\
(f) $($Barbadori$,$ Salviati$)$ and $($Castellani$,$ Pazzi$);$\\
(g) $($Ginori$,$ Aciaiuoli$,$ Pazzi$,$ Lamberteschi$)$ and $($Albizzi$,$ Medici$,$ Salviati$,$ Guadagni$).$
\end{lemma}

\begin{proof}(a) Observe that Tornabuoni and Albizzi have three neighbors each, and they share two neighbors. Therefore, by S, the relation between them is the same as the relation between the remaining neighbors, Ridolfi and Ginori.
(b)~Bischeri and Peruzzi are adjacent and have a common neighbor Strozzi; in addition, Bischeri has a neighbor Guadagni, while Peruzzi has a neighbor Castellani.
\x{It is easy to verify that by}Due to S, the relation between Bischeri and Peruzzi coincides with that between Guadagni and Castellani. Indeed, it is easy to see that the edge \{Bischeri, Peruzzi\} cannot fix the violation of Self-consistency that may occur in the absence of this edge. This completes the proof of~(b). The remaining assertions are proved similarly.
%\qed
\end{proof}

%\medskip
\smallskip
The following proposition demonstrates that Lemma~\ref{l:5} can be quite useful in proving the violation of\x{ that certain measures violate} Self-consistency.

\begin{proposition}\label{p:9} %9.
%Centrality measures
\Walk$(K_{ii}),$ \Communicability$(K_{ii}),$ \Closeness{$($\xz\Forest$),$} \Closeness$($\xz\Heat$),$ \Closeness$^*(${$\log$\Walk}$),$ \Closeness$^*(${$\log$\Communicability}$),$ \Closeness$^*(${$\log$\Forest}$),$ and \Closeness$^*(${$\log$\Heat}$)$ centralities
with parameter $t=1$ for the Forest, Heat, and Communicability kernels and $t=(\rho(A)+1)^{-1}$ for the Walk kernel
violate \Axiom~S on the graph of Fig.~\ref{f:Floren}.
\end{proposition}

\begin{proof}For the graph in Fig.~\ref{f:Floren}, \Walk$(K_{ii})$ and \Communicability$(K_{ii})$ provide a centrality ranking in which Peruzzi $\succ$ Bischeri, but Guadagni  $\succ$ Castellani. Thus, by part~(b) of Lemma~\ref{l:5}, these measures violate Self-consistency.
Measures \Closeness$($\xz\Forest$),$ \Closeness$^*(${$\log$\Walk}$),$ \Closeness$^*(${$\log$\Communicability}$),$ and \Closeness$^*(${$\log$\Heat}$)$ provide rankings in which Ridolfi $\succ$ Tornabuoni, but Guadagni  $\succ$ Strozzi. Thus, by part~(e) of Lemma~\ref{l:5}, these measures violate Self-consistency.
Measures \Closeness$($\xz\Heat$),$ and \Closeness$^*(${$\log$\Forest}$)$ provide rankings in which Castellani $\succ$ Peruzzi, but Bischeri  $\succ$ Barbadori. Thus, by part~(d) of Lemma~\ref{l:5}, these measures violate Self-consistency.
%\qed
\end{proof}

\smallskip
Note that Table~1 in \cite{CheGub20How-} provides additional information on the violation of \Axiom~S by\x{ some other} centrality measures.

\section{Core intuition behind centrality and {\em PageRank} in its light} % \x{ underlying}-20%
\label{s:intuit}

\x{A perfect }The best example of a ``central'' node is the center of a star of order more than~$2.$
Formally, a star of order $n$ is a graph with one node (the \emph{center}) having degree $n-1$ and $n-1$ nodes\x{ are leaves} having degree~$1.$
The edges of a star are sometimes called {\em rays}.
\medskip

According to Freeman \cite{freeman1978centrality}, ``one general intuitive theme seems to have run through all the earlier thinking about point centrality in social networks: the point at the center of a star [...] is the most central possible position''.
%\smallskip

\begin{definition}\label{d:StaRC}
We say that a centrality measure on a star $G$ with two or more rays\x{ $n>2$ nodes} satisfies the {\em star condition\/} if it assigns maximum centrality exclusively to the center of this star.
\end{definition}

An example of a centrality measure that violates the star condition is given in~\cite[Section~1]{CheGub20How-}.

Self-consistency is a rather strong axiom, however, as was noted in Section~\ref{s:axioms}\x{C}, it is not self-sufficient\x{ comprehensive} and requires some accompanying axioms.
One of its \x{features}peculiarities is that it only applies to nodes of the same degree.
Therefore, it does not imply\x{ that the center of a star has the maximum centrality} the star condition.
As distinct from it, the Bridge axiom\x{ ensures the discussed property} implies this condition.

\begin{proposition}\label{p:BSt}
On a star with two or more rays$,$ any centrality measure that satisfies \Axiom~B also meets\x{ satisfies} the star condition.
\x{\hl{Check papers.}}
\end{proposition}

\begin{proof}
This is true because each ray of a star is a bridge, and among the components formed after its removal, the component containing a leaf is smaller than that containing the center.
%\qed
\end{proof}

\medskip
%\bigskip
However, \Axiom~B does not imply that the centrality of all leaves of a star is the same, which is immediate from Self-consistency (or from~\x{\Axiom~}\Ee\x{, as the leaves are equivalent}).
\medskip

Roy and Tredan \cite{RoyTredan10SharpCentr}, trying to capture the intuition behind\x{ underlying} the concept of centrality, claim that for
a path graph with nodes $1\cdc n$, where each node $u$ such that ${1<u<n}$ is linked to $u-1$ and $u+1,$ it is (converting to our notation) ``hard to imagine a centrality $f$ such that, given a node $u$ $(u\ne\frac{n+1}2),$ we have $f(u)\not\in[f(u-1),f(u+1)]$''.
\x{\hl{Check~PR.}}

\begin{definition}\label{d:RT}
Let $G$ be a path graph where each node $u$ such that ${1<u<n}$ is linked to $u-1$ and $u+1$. A centrality measure $f$ on $G$ is said to satisfy the
\begin{itemize}
\item
{\em Roy-Tredan $($RT\/$)$ condition\/} if for any node $u,$ $u\not\in\left\{1,\frac{n+1}2,n\right\}\To f(u)\in[f(u-1),f(u+1)];$
\item {\em \pathcondition\/} if the centrality of a node strictly increases with increasing \x{ the}shortest path distance from the nearest leaf.
\end{itemize}
\end{definition}
%\smallskip

Clearly, under~\x{\Axiom~}\Ee\ and $n>2$\x{C} the \pathcondition\ is stronger than the RT condition.
Proposition~\ref{p:BPath} states that the \pathcondition\ follows from \Axiom{s}~B and~\Ee.

\begin{proposition}\label{p:BPath} %2.
For a path graph$,$ any centrality measure that satisfies \Axiom{s}~\Ee\ and B meets\x{ also satisfies} the \pathcondition.
\end{proposition}

\begin{proof}
Let $f$ satisfy \Axiom{s}~\Ee\ and B.
Consider the path graph 1---2---$\xxy\cdots$---$n,$ where $V=\{1\cdc n\}$ and ``---'' denotes an edge. Let $1\le u=v-1<n.$
Suppose first that $v\le\frac{n+1}2.$
Then $\{u,v\}\in E$ is a bridge and by \Axiom~B, $f(u)<f(v)$ since $|V_u|<|V_v|.$ Hence for such $u$ and $v,$ the centrality strictly increases with increasing distance from the nearest leaf~$1.$
The case of $v>\frac{n+1}2$ is considered similarly.
Finally, if $u<\frac{n+1}2$ and $u-1=n-w,$ i.e., $u$ and $w$ have the same distance from the nearest leaves, then $u\sim w$ and by~\x{\Axiom~}\Ee, $f(u)=f(w).$ This implies the \pathcondition.
%\qed
\end{proof}

\medskip
It is all the more remarkable that \PageRank, one of the most popular\footnote{\x{According}To quote\x{C} \cite[\hl{p.\,12530}]{Tu18PageRank}, ``PageRank centrality is probably the most well-known and frequently used measure''; see also \cite{Avrachenkov22Book}.} centrality measures, according to Roy and Tredan is ``hard to imagine'' because it violates the RT condition.

\begin{proposition}\label{p:BP_PR}
On the path graph $1$---$2$---$3$---$4$---$5,$ \PageRank\ centrality $\fPRa$ with any $\a\in(0,1)$ violates the RT and the \pathcondition{s}. Namely$,$ $\fPRa(2)>\max\{\fPRa(1),\fPRa(3)\}.$
\end{proposition}

\begin{proof}
For the path graph 1---2---3---4---5, let us search the Perron eigenvector that solves~\eqref{e:PR} in the form $\bm x=(a,b,c,b,a)^T$ (as \PageRank\ satisfies \Ee) with $a+b+c+b+a=1$. \x{the sum of the components equal to~$1$.}
Then equations~\eqref{e:PR} have the form %By~\eqref{e:PR} we have
\begin{eqnarray}
a&=&        0.5\a b + 0.2(1-\a);\nonumber\\
b&=&\a(   a + 0.5c) + 0.2(1-\a);\label{e:PRn}\\
c&=&\a(0.5b + 0.5b) + 0.2(1-\a).\nonumber    %\a b
\end{eqnarray}

Combining them we obtain $(1+0.5\a)(b-a)=0.5\a(a+c),$ whence $b>a,$ i.e., $\fPRa(2)>\fPRa(1),$ and
\eq{e:b-c}{(1+0.5\a)(b-c)=\a(a-0.5b)=0.5\a(1-\a)(0.4-b).}
If $c\ge b,$ then by \eqref{e:b-c} $b\ge0.4$ and $2b+c>1.2$, which contradicts $2b+c+2a=1,$ consequently, $b>c,$ i.e., $\fPRa(2)>\fPRa(3).$
Thus, all \PageRank\ centralities violate the RT and the \pathcondition{s} on this path graph.
\end{proof}

\medskip
Node 3 of the 1---2---3---4---5 path can be considered as its center. It follows from the proof of Proposition~\ref{p:BP_PR} that \PageRank\ never assigns the maximum centrality to this center\x{node}.
It can be shown that while 1---2---3---4---5 is the minimal path on which \PageRank\ violates the RT condition, this centrality also violates it and the \pathcondition\ on the\x{C} path graphs with $n>5.$%larger numbers of nodes.

\begin{corollary}\label{c:PRB} %3.
\PageRank\ centrality with any $\a\in(0,1)$ violates \Axiom~B on the path graph of order~$5$.
\end{corollary}

\begin{proof}
By Proposition~\ref{p:BP_PR}, on the path graph 1---2---3---4---5, $\fPRa(2)>\fPRa(3)$ for any $\a\in(0,1)$.
This violates B, since $\{2,3\}$ is a bridge and, in the notation of the Bridge axiom, $|V_2|<|V_3|.$
\end{proof}

\smallskip
\begin{proposition}\label{p:PR} %3.
\PageRank\ centrality with any $\a\in(0,1)$ violates \Axiom~S on the path graph of order~$5$.
\end{proposition}

\begin{proof}
In the path graph 1---2---3---4---5\x{ denoted by $H_5$}, node 2 has neighbors 1 and 3, 3 has neighbors 2 and 4; by Proposition~\ref{p:BP_PR}, for any $\a\in(0,1),$ $\fPRa(2)>\fPRa(1)$ and $\fPRa(4)=\fPRa(2)>\fPRa(3),$ i.e., for some bijection between $N_3$ and $N_2$, all neighbors of 3 have higher centrality values than the corresponding neighbors of~2. Hence \Axiom~S requires $\fPRa(3)>\fPRa(2),$ which is false by Proposition~\ref{p:BP_PR}. Therefore, all \PageRank\ centralities violate \Axiom~S on the path graph of order~$5$.
%\qed
\end{proof}

\medskip
It will be shown in Corollary~\ref{c:PRT} that \PageRank\ centrality also violates one of monotonicity axioms.
In Section~\ref{s:Discuss}, we discuss the specifics of \PageRank\x{ in detail} and point out some types of networks for which such a centrality\x{ measures} may be useful.

\begin{remark}\label{r:KernelRT}
{Among the kernel-based centralities, there are also measures that violate the RT or} \pathcondition.
For example, on the 1---2---3---4---5 path graph,
\Eccentricity(\Communicability) sets $f(2)>f(1)=f(3)$ when $t$ is sufficiently large and $f(1)=f(2)<f(3)$ when $t$ is smaller (with $f(1)=f(2)=f(3)$ on the boundary with $t\approx 4.017$).
\Closeness(\Communicability) sets $f(2)>f(1)>f(3)$ when $t>t_1\approx7.296$, $f(2)>f(3)>f(1)$ for smaller values of $t$, and $f(3)>f(2)>f(1)$ when $3.631\approx t_0>t>0.$

Similarly, \Eccentricity(\Walk) sets $f(2)>f(1)=f(3)$ when $\frac1{\sqrt{3}}=(\rho(A))^{-1}>t>t_2\approx0.5486$ and $f(1)=f(2)<f(3)$ when $t<t_2$ (with $f(1)=f(2)=f(3)$ when $t=t_2$).
\Closeness(\Walk) sets $f(2)>f(1)>f(3)$ when $(\rho(A))^{-1}>t>t_1\approx0.5755$, $f(2)>f(3)>f(1)$ when $t_1>t>t_0\approx0.5345$, and $f(3)>f(2)>f(1)$ when $t_0>t>0.$

Such peculiarities (related to horseshoe-like or even tweezers-like curvilinear space representations of the 1---2---3---4---5 path graph) are not characteristic\x{ typical} of kernel-based measures and are not properties of the $\log$\Communicability\ or $\log$\Walk\ \hl{based} \Eccentricity\ or \Closeness\ measures. {This partly explains the fact that $\log$-measures outperform plain ones in cluster analysis} \cite{ivashkin2016logarithmic,IvChe21}.
\end{remark}

Note that if the Self-consistency or Bridge axiom is considered as a mandatory\x{ n indispensable} property of a centrality, then this can drastically reduce the set of candidate measures (see \cite[Figures~3, 7, and~8]{CheGub20How-}).

\section{Contribution of monotonicity axioms}%{Combinations with}
\label{s:Monot}

In this section, we focus on edge-monotonicity conditions, which, along with\x{as well as} the Self-consistency and Bridge axioms, belongs to the class of positive responsiveness axioms.
We prove that together with \Axiom{s}~\Ee\ and, in one case, S, they imply the star and path centripetal conditions and contradict \Axiom~B, while \PageRank\ violates {not only}\x{C} \Axiom{s}~B ans S, but also\x{C} Transit monotonicity.

The edge-monotonicity axioms\x{ of this section} involve two graphs: an original graph $G_0$ and a graph $G$ obtained from $G_0$ by adding an extra edge (or edges). These axioms restrict the set of {\em universal centrality measures\/} $f_G(u)$ that evaluate the nodes of {\em all\/} connected graphs $G$ with $n>1$. %The word ``universal'' in the formulations of Propositions~\ref{p:1} to \ref{p:incomp} is implied, not explicit.

\begin{axiom}[M: Monotonicity]{ Suppose that $u,v\in V(G_0),$ $f_{G_0}(u)\le f_{G_0}(v),$ $u\ne v,$ and
%$G^+$ is obtained from $G$ by adding an edge $\{v,w\},$ where $w\ne u.$
$G=G_0\cup G^+\ne G_0,$ where $V(G^+)=\{v,w\},$ $E(G^+)=\{\{v,w\}\},$ and $w\ne u.$
Then ${f_G(u)<f_G(v)}$.}
\end{axiom}

According to Monotonicity, if $v$ is at least as central as $u$\x{ is no more central than $v$} and a new edge not incident to $u$ is attached to $v,$ then $v$ becomes or remains more central than~$u.$
A stronger version of \Axiom~M that also states that the relation between the centralities of $u$ and $v$ remains the same after the addition of edge $\{u,v\}$ makes sense as well, but we do not consider it in this paper.

Similar axioms called Adding rank monotonicity and Strict rank monotonicity have been proposed in \cite{csato2017measuring,Boldi22Monotonicity-} and \cite{ChienDwork04link,boldi2017rank} (for directed graphs)\x{, respectively}. Item~$1.2$ of Dynamic monotonicity in~\cite{Che94} is the corresponding condition for directed graphs representing paired comparisons.

Monotonicity together with~\x{\Axiom~}\Ee\ imply the star condition.
\begin{proposition}\label{p:1} %1.
Any universal centrality measure that satisfies \Axiom{s}~\Ee\ and M meets the star condition and assigns the same centrality to all star leaves\x{ on any star with at least two rays}.
\end{proposition}

\begin{proof} By \Ee, the centrality of the two nodes of a 1-ray star is the same. By M, adding a\x{ one more} node adjacent to the ``center'' of the 1-ray star makes the centrality of the center greater than the same centrality of the leaves, and attaching additional leaves preserves this.
%\qed
\end{proof}

\medskip
%\subsection{Paths} %4.2
Transit monotonicity is a natural and essential strengthening of~M.

\begin{axiom}[T: Transit monotonicity]{ If\x{ $u,v\in V(G_0),$} $f_{G_0}(u)\le f_{G_0}(v),$ $u\ne v,$ $G=G_0\cup G^+\ne G_0$, and any path in $G\/$ from a node of $G^+$ to $u$ contains $v,$ then $f_G(u)<f_G(v).$}
\end{axiom}

According to \Axiom~T, if $v$ is at least as central as $u$ and $v$ is a cutpoint between new edges and $u$, then $v$ becomes or remains more central than $u$ after the addition of these edges.

\smallskip
Together with~\x{\Axiom~}\Ee, Transit monotonicity implies the \pathcondition.

\begin{proposition}\label{p:2} %2.
\x{For a path graph$,$ }Any universal centrality measure that satisfies \Axiom{s}~\Ee\ and T meets\x{ also satisfies} the \pathcondition.
\end{proposition}

\begin{proof}\x{ In this proof, we omit the subscript $G$ of a universal centrality measure $f_G,$ since each graph is described explicitly without introducing a special notation.} Let $f_G$ be a universal centrality measure satisfying \Axiom{s}~\Ee\ and~T.
By \Ee, the conclusion of Proposition~\ref{p:2} holds for the 1---2 path graph. To proceed by induction,
assume that it is true for a path graph 1---$\xxy\cdots$---$2k$, $k\in\N$, denoted by~$H_{2k}$.
Then $f_{H_{2k}}(i)\le f_{H_{2k}}(i+1)$ for all $i\in\{1\cdc k\}.$
Attaching a new node $2k+1$ and the edge $\{2k,2k+1\}$ yields the path graph 1---$\xxy\cdots$---$(2k+1)$ denoted by $H_{2k+1}$.
Since any path in $H_{2k+1}$ from $2k+1$ to $i\in\{1\cdc k\}$ contains $i+1,$ \Axiom~T implies $f_{H_{2k+1}}(i)<f_{H_{2k+1}}(i+1).$
Therefore, the centrality of the nodes $i\in\{1\cdc k+1\}$ of $H_{2k+1}$ strictly increases with the increase of the shortest path distance from the nearest leaf (in this case, leaf~$1$).
%By the assumption and \Axiom~T, the conclusion of Proposition~\ref{p:2} holds for the nodes $1,2\cdc (k+1).$
The same holds for all nodes\x{ $i\in\{k+2\cdc 2k+1\}$} due to the equivalence $i\sim (2k+2-i)$ for all $i\in\{1\cdc k\}$ and~\x{\Axiom~}\Ee.
Thus, the conclusion of Proposition~\ref{p:2} is true for $H_{2k+1}$.\x{ the 1---$\xxy\cdots$---$(2k+1)$ graph.} Adding node $2k+2$ and edge $\{2k+1,2k+2\}$ to $H_{2k+1}$, we similarly derive that this conclusion also holds for the resulting 1---$\xxy\cdots$---$(2k+2)$\x{ path} graph. This completes the proof.
%\qed
\end{proof}

\medskip
As a consequence of Proposition~\ref{p:2}, we obtain that any \PageRank\ centrality violates \Axiom~T.

\begin{corollary}\label{c:PRT} %3.
Any universal centrality measure that satisfies \Ee\ and coincides with a \PageRank\ centrality on the path graph of order~$5$ violates \Axiom~T.
\end{corollary}

\begin{proof}
By Proposition~\ref{p:2}, any universal centrality measure that satisfies \Axiom{s}~\Ee\ and T meets the \pathcondition\x{ on the path graph of order~$5$}. Hence it cannot coincide with a \PageRank\ centrality on the path graph of order~$5$, since the latter centrality violates the \pathcondition\ on this graph for any $\a\in(0,1)$ by Proposition~\ref{p:BP_PR}.
%\qed
\end{proof}

\medskip
On some other peculiarities of\x{ the} \PageRank\ centralities, see~\cite[Section~1]{CheGub20How-}.

According to \cite{Boldi22Monotonicity-}\x{[Theorems~12--15]}, for each graph $G_0$ and large enough $\a\in(0,1)$, {\em no\/} universal centrality measure coinciding with $\fPRa$ on $G_0$ along with all graphs $G$ obtained from $G_0$ as in \Axiom~M violates M\x{this axiom} on any pair $(G_0,G)$. However,\x{On the other hand, by \cite[Theorems~14--15]{Boldi22Monotonicity-},} no universal\x{ centrality} measure that reduces to $\fPRa$ with a fixed $\a$ on {\em all\/} connected graphs with $n>1$ satisfies \Axiom~M. For \PageRank\ universal measures with varying damping factor \x{C}$\a$, prospective\x{the} results depend on the specific function $\a(G).$
The situation with \Katzc\ and \GeneralizedDegree\ centralities is similar with the difference that here $\a$ must be small enough for M to hold \cite{Boldi22Monotonicity-,csato2017measuring}. %[Theorems~9--11]
The \Eigenvector\ centrality violates M \cite{Boldi22Monotonicity-}.

\smallskip
We conclude by showing that under \Equivalence, the conjunction of M and S is incompatible with \Axiom~B, and so is~T.

\begin{proposition}
\label{p:incomp}
If a universal centrality measure satisfies \Axiom{s}~\Ee$,$ M$,$ and S or \Axiom{s}~\Ee\ and T$,$ then it violates \Axiom~B.
\end{proposition}

\begin{proof}
Let a universal centrality measure $f_G$ satisfy \Axiom{s}~\Ee\ and~B.
For the graph $G$ in Fig.~\ref{f:Counter}a, $f_G(4)=f_G(3)$ by~B. On the other hand, for the graph $G_0$ in Fig.~\ref{f:Counter}b, $f_{G_0}(4)=f_{G_0}(3)$ by~\Ee\ or~B. Observe that $G=G_0\cup G^+,$ where $V(G^+)=\{0,1\}$ and $E(G^+)=\{\{0,1\}\}.$
\begin{figure}[!ht]
\begin{center}
{\small (a)\hspace{15em}(b)}

\includegraphics[height=5.7em]{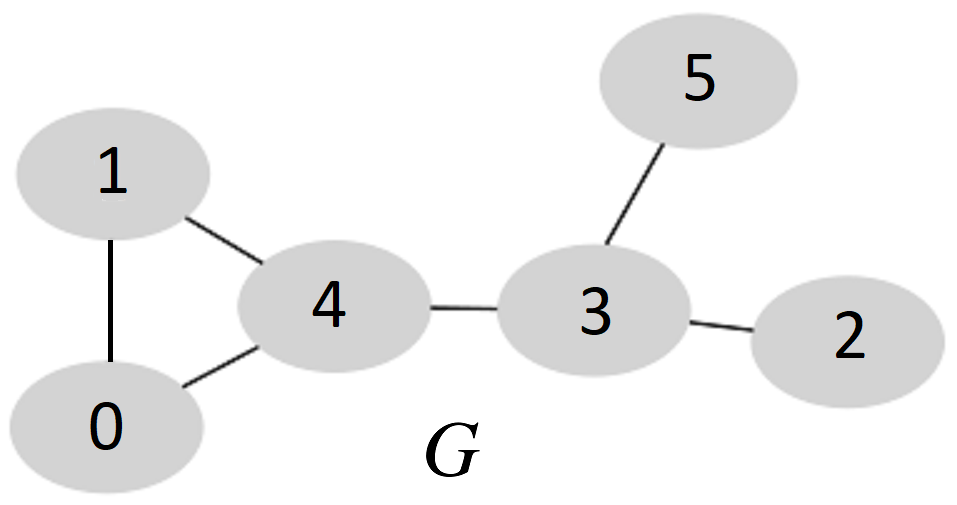}\qquad\qquad   %6em
\includegraphics[height=5.7em]{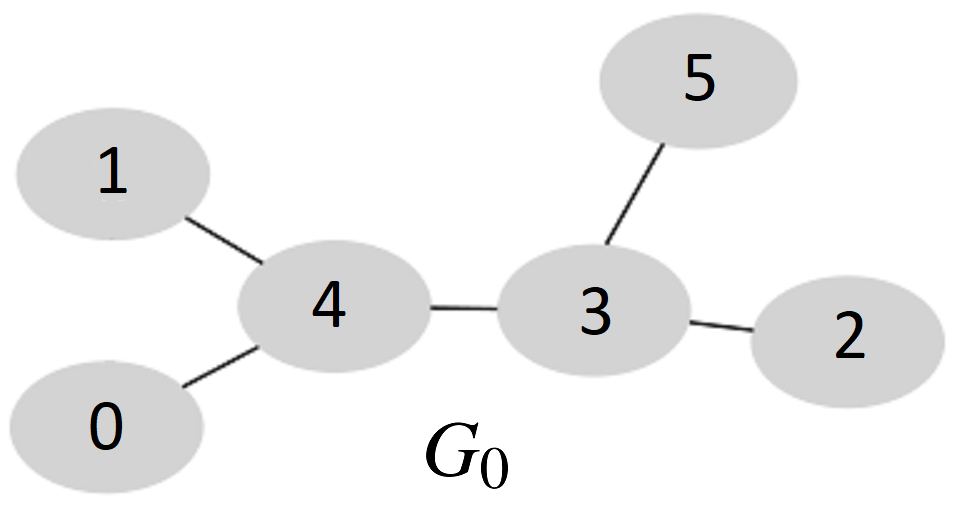}
\caption{(a)~The graph $G$ and (b)~its subgraph $G_0$ used in the proof that \Ee\&M\&S as well as {\Ee\&T} are incompatible with \Axiom~B (Proposition~\ref{p:incomp}).\label{f:Counter}}
\end{center}
\end{figure}

Assume that $f_G$ also satisfies \Axiom{s}~M and~S. By \Ee\ and M, $f_G(0)=f_G(1)>f_G(2)=f_G(5).$ Therefore, by S,\x{ we have} $f_G(4)>f_G(3),$ a contradiction.

Now assume that, instead of  M\&S, $f_G$ satisfies \Axiom~T. Since $f_{G_0}(4)=f_{G_0}(3)$ and all paths in $G$ from 0 or 1 to 3 contain 4, T implies $f_G(4)>f_G(3)$, a contradiction.

Thus, the conjunctions \Ee\&M\&S\&B and \Ee\&T\&B are both false.
%\qed
\end{proof}

%\smallskip
\begin{remark}\label{r:BnoT}
The reason why \Axiom{s}~B and T are incompatible under \Equivalence\ is easy to  discern.
If $\{u,v\}$ is a bridge in $G$ and $|V_u|=|V_v|$, then B implies $f(u)=f(v).$ However, if the restriction of $G$ to $V_u$ is isomorphic to a proper subgraph of the restriction of $G$ to $V_v$ with $u$ corresponding to $v$, then T requires $f(u)<f(v).$
The logic of \Axiom~S is similar to that of T in terms of the transfer of influences, however, S is not ``grounded'' as it does not imply any\x{ impact} influence of local ``graph density'' on centrality. In the conjunction M\&S, axiom M provides a kind of ``grounding'' leading to a contradiction with~B.
\end{remark}

\section{Discussion}\label{s:Discus} %{Conclusion}  (and future work)
\label{s:Discuss}

Each point centrality index measures some structural capital of the nodes.
According to the Bridge axiom, one end-node of a bridge is more central than the other if and only if the\x{ deletion} removal of the bridge leaves the first one in a greater (in terms of the number of nodes) component. In this sense, the corresponding capital is {\em node}-based: it does not depend on the density of the components. Self-consistency states that the capital of a node increases with the capital of its neighbors. By the Monotonicity axiom, edges incident to a node contribute to its capital, so the capital measured by centrality indices satisfying Monotonicity is locally {\em edge}-based. The conjunction of Self-consistency and Monotonicity spreads the influence of edges throughout the graph, making it global. As a result, this conjunction turns out to be incompatible (under \Equivalence) with the node-based Bridge axiom (Proposition~\ref{p:incomp}). By the same proposition, the Bridge axiom is incompatible with the Transit monotonicity axiom, which postulates the edge nature of the capital globally.

According to Lemma~\ref{l:CutBridge}, any \Closeness\ centrality whose corresponding distance is cutpoint additive satisfies the Bridge axiom.
A universal centrality measure satisfies Self-consistency if and only if \hl{all centrality vectors this measure assigns to connected graphs have} \monotonic\ neighborhood representations (Lemma~\ref{l:monSC}).

In this paper, we have paid some attention to the properties\x{ related to the main topic} of \PageRank\ centrality. It turns out that this index\x{ measure} violates most of the conditions under consideration and even has a property that, according to some authors, ``is hard to imagine'' for a measure of centrality.
The feature responsible for this is the combination of stochastic normalization performed in \eqref{e:PR} by multiplying $A^T$ by $(\diag(A\bm1))^{-1}$ and uniform ``teleportation''.

This can be seen\x{explained}\x{C} by returning to the path graph $1$---$2$---$3$---$4$---$5$ appeared in Proposition~\ref{p:BP_PR}.
In the second equation of \eqref{e:PRn}, centrality $a$ of node $1$ contributes to the index value $b$ of node $2$ with the maximum possible weight $1\ccdot\a$. This is because the stochastic normalization of the\x{ first} column of $A^T$ corresponding to node $1$ does not change\x{ preserves} its entries, since there is only one $1$ in this column. The weight of node's $3$ centrality in the same equation is $0.5\ccdot\a$ because this node has two edges. Comparing to this, the centrality of node $3$ also has two contributions from the neighbors, but both with weights of $0.5\ccdot\a$, since neighboring nodes $2$ and $4$ each have two edges. Thus, node $2$ has an ``advantage'' over node $3$ in the contribution weights\x{: $1.5\ccdot\a$ against $1\ccdot\a$}.

Substituting\x{C} $\a=1$ into \eqref{e:PR} results\x{ \PageRank\ centrality reduces to} in \Seeley\ centrality, which is proportional to the node degree \hl{in the undirected case}. In this case, the centrality of $3$ (or $2$) is twice that of $1$, which according to \eqref{e:PRn} arithmetically provides the same centrality for $2$ and $3$ with the above advantage of $2$ in the contribution weights. However, if $\a<1,$ then the uniform teleportation brings the centralities of nodes with different degrees closer together so that the centrality of $1$ increases relative to those of $2$ and $3$. As follows from the proof of Proposition~\ref{p:BP_PR}, this results in $2$ getting ahead of $3$ due to its advantage in contribution weights.

The source of the above advantage is that $1$ is a leaf. Thus, we conclude that among nodes of the same degree, \PageRank\ rewards\x{C} those that are connected to nodes of low degree, since the contribution weights are the reciprocals of neighbors' degrees. This is contrary to the logic of Self-consistency provided that the centrality is positively correlated with the node degree.
Measures using non-normalized contribution weights, including  \GeneralizedDegree, \Eigenvector, and \Katzc-\Bonacich, have \monotonic\ neighborhood representations and due to Lemma~\ref{l:monSC} satisfy Self-consistency.
\begin{figure}[!ht]
\begin{center}
{\small (a)\hspace{15em}(b)}

\includegraphics[height=5.94em]{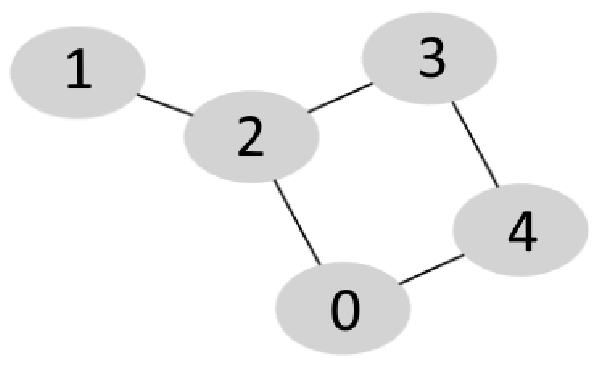}\qquad\qquad   %6em
\includegraphics[height=8.69em]{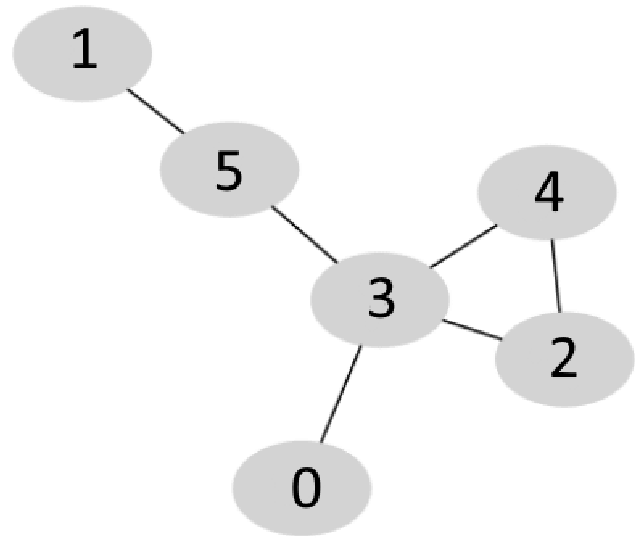}
%\\\includegraphics[height=5.94em]{G2a.png}\qquad\qquad   %6em
%\includegraphics[height=8.69em]{G7a.png}
\caption{~For these graphs, (a) $\fPRa(4)>\fPRa(0)$; (b)~$\fPRa(1)>\fPRa(0)$.\label{f:PR2}}
\end{center}
\end{figure}

Consider two more examples: for the graphs in Fig.~\ref{f:PR2}, with any $\a\in(0,1)$ (a)~the \PageRank\ centrality of node $4$ is higher than that of node $0$; (b)~$\fPRa(1)>\fPRa(0)$ due to the fact that some neighbors of these nodes $4$ and $1$ (in subfigures (a) and (b), respectively) have\x{ (partially)} lower degrees than the neighbors of nodes labeled~$0$.

Centrality measures of this kind can be useful for some types of undirected friendship networks. If a person has $k$ friends for whom she is the only friend, then the responsibility of this person is higher than that of a person with $k$ friends, each of whom has many friends.
Similarly, if each connection in a network is a mutual assistance agreement, then fixing the number of personal agreements, the agent can attract more partner resources if the partners do not have too many other obligations. Thus, \PageRank\ may be an appropriate index to measure such responsibility or resource availability.

The axioms of Self-consistency and Bridge are quite\x{ fairly} strong, so the adoption of either\x{ each} of them drastically\x{ dramatically} reduces the set of centrality measures under consideration.
This fact is used in the application of\x{C} the culling method \cite{CheGub20How-} designed to select the most appropriate centrality measures.
This method consists in compiling and completing a survey based on a decision tree like that in Fig.~\ref{f:Mtree} that allows the user to find a measure that matches\x{ is consistent with} their idea\x{ underlying concept} of centrality.
In the framework of this method, adopting a certain axiom results in compiling a shorter survey on the set of\x{ reduced to the} measures that satisfy this axiom. In \cite{CheGub20How-}, the surveys reduced to the measures satisfying the Self-consistency or Bridge axioms are shown in Figures~7 and~8, respectively.

\section*{Acknowledgments}
This work was supported by the European Union (ERC, GENERALIZATION, 101039692). Views and opinions expressed are however those of the author only and do not necessarily reflect those of the European Union or the European Research Council Executive Agency. Neither the European Union nor the granting authority can be held responsible for them.

The author thanks Anna Khmelnitskaya, Dmitry Gubanov, and Konstantin Avrachenkov for helpful discussions and two anonymous reviewers for their careful comments and thoughtful suggestions.

%\small
\bibliographystyle{comnet} %{elsarticle-num}
\bibliography{centrality}
\end{document}